\documentclass[conference]{IEEEtran}
\IEEEoverridecommandlockouts
\def\BibTeX{{\rm B\kern-.05em{\sc i\kern-.025em b}\kern-.08em
    T\kern-.1667em\lower.7ex\hbox{E}\kern-.125emX}}

\usepackage{tikz}
\usepackage{tikz-qtree}
\usepackage{adjustbox}
\usetikzlibrary{calc}
\usetikzlibrary{decorations.pathmorphing}
\usepackage{calc}
\usepackage{amsmath}
\usepackage{multirow}
\usepackage{array}
\usepackage{soul}
\usepackage{mdframed}
\usepackage{xspace}
\usepackage{booktabs}
\usepackage{cleveref}
\usepackage{subcaption}
\usepackage{graphicx}
\usepackage[printonlyused]{acronym}
\newtheorem{definition}{Definition}

\newcommand{\wx}{\hskip 4pt}
\soulregister{\ac}{7}

\newcommand{\setto}{\gets}

\newcommand{\createproof}{\textrm{CreateProof}\xspace}
\newcommand{\findmissing}{\textrm{FindMissingChunks}\xspace}

\newcommand{\chunkstoreprover}{\mathcal{CP}\xspace}
\newcommand{\chunkstoreverifier}{\mathcal{CV}\xspace}

\newcommand{\argproof}{\textsl{proof}\xspace}

\newcommand{\argaddr}{\textsl{id}\xspace}
\newcommand{\msgprove}     {\textsc{Prove}\xspace}
\newcommand{\msgselect}    {\textsc{Select}\xspace}
\newcommand{\msgupload}    {\textsc{Upload}\xspace}
\newcommand{\msguploaddone}{\textsc{UploadDone}\xspace}
\newcommand{\msgnewproof}  {\textsc{NewProof}\xspace}
\newcommand{\concat}{\mathbin\Vert}

\newcommand{\hash}{\ensuremath{\mathrm{H}}\xspace}
\newcommand{\funmphf}{\ensuremath{\mathrm{MPHF}}\xspace}
\newcommand{\findchunkproof}{\ensuremath{\mathrm{Find}}\xspace}

\newcommand{\vardata}{\textsl{chunk}\xspace}
\newcommand{\varbaseproof}{\textsl{chunkProofs}\xspace}

\newcommand{\varpublick}{\textsl{nonce}\xspace}
\newcommand{\varstart}{\textsl{start}\xspace}
\newcommand{\varend}{\textsl{end}\xspace}

\newcommand{\varval}{\textsl{idx}\xspace}
\newcommand{\varmapvalid}{\textsl{index-id}\xspace}
\newcommand{\varmapidproof}{\textsl{id-proof}\xspace}
\newcommand{\varmphf}{\textsl{mphf}\xspace}

\newcommand{\varinterval}{\textsl{i}\xspace}
\newcommand{\varmissing}{\textsl{missing}\xspace}
\newcommand{\peershort}{\textsl{p}\xspace}

\newcommand{\varchunkproofs}{\textsl{cp-id}\xspace}

\newcommand{\varchunkproof}{\textsl{cp}\xspace}
\newcommand{\varchunkid}{\textsl{chunkid}\xspace}
\newcommand{\varexpectedvals}{\textsl{mfc}\xspace}
\newcommand{\varnewproof}{\textsl{collision}\xspace}
\newcommand{\varproofsize}{\textsl{size}\xspace}

\usepackage{amsmath}
\usepackage{algorithm}
\usepackage[noend]{algpseudocode}
\usepackage{enumitem}

\algnewcommand\swt[1]{\State\textbf{switch} #1:}

\algsetblockdefx[MyFunc]{MyFunc}{EMyFunc}
{}{.8em}
[2]{{\bf func} #1(#2)}
{EndFunc}
\algnotext{EMyFunc}

\algnewcommand\algorithmicforeach{\textbf{for each}}
\algdef{S}[FOR]{ForEach}[1]{\algorithmicforeach\ #1\ \algorithmicdo}

\algdef{SE}[RECEIVE]{Receive}{EndReceive}[1]{\textbf{upon receive}\ #1\ \algorithmicdo}{\algorithmicend\ \textbf{receive}}%
\algtext*{EndReceive}

\algrenewcommand{\algorithmiccomment}[1]{{\algcommentfontsize\hfill$\triangleright$~#1}}

\algnewcommand{\algorithmicand}{\textbf{ and }}
\algnewcommand{\algorithmicor}{\textbf{ or }}

\newcommand{\algfontsize}{\footnotesize}
\newcommand{\algcommentfontsize}{}

\newcommand{\true}{\textsl{true}\xspace}
\newcommand{\false}{\textsl{false}\xspace}

\newcommand{\reply}{\textbf{reply}\xspace}
\newcommand{\return}{\textbf{return}\xspace}
\newcommand{\send}{\textbf{send}\xspace}
\newcommand{\new}{\textbf{new}\xspace}

\newcolumntype{L}[1]{>{\raggedright\arraybackslash}p{\dimexpr#1\columnwidth-2\tabcolsep\relax}}
\newcolumntype{C}[1]{>{\centering\arraybackslash}p{\dimexpr#1\columnwidth-2\tabcolsep\relax}}
\newcolumntype{R}[1]{>{\raggedleft\arraybackslash}m{\dimexpr#1\columnwidth-2\tabcolsep\relax}}

\newcommand{\mphf}{\ac{MPHF}\xspace}
\acrodef{MPHF}{Minimal Perfect Hash Function}
\newcommand{\protocolname}{\ac{SNIPS}\xspace}
\acrodef{SNIPS}{succinct proof of storage}
\acrodef{PoS}{Proof of Storage}
\acrodef{IBF}{invertible Bloom filter}

\newcommand*\rightbracket{\textrm{]}}
\newcommand*\leftbracket{\textrm{[}}

\newcommand{\swarmbee}{Swarm~Bee\xspace}
\newcommand{\snipskeywords}{data synchronization, proof of storage, decentralized storage system, maintenance, redundancy, data upkeep, set reconciliation \xspace}
\newcommand{\pullsync}{Pullsync\xspace}
\newcommand{\swarmbeeversion}{v1.9.0\xspace}

\begin{document}

\title{SNIPS: Succinct Proof of Storage for Efficient Data Synchronization in Decentralized Storage Systems}

\author{
\IEEEauthorblockN{Racin Nygaard}
\IEEEauthorblockA{\textit{University of Stavanger}} \and
\IEEEauthorblockN{Hein Meling}
\IEEEauthorblockA{\textit{University of Stavanger}}
}
\maketitle
\thispagestyle{plain}
\pagestyle{plain}

\begin{abstract}
Data synchronization in decentralized storage systems is essential to guarantee sufficient redundancy to prevent data loss.
We present \emph{\acs{SNIPS}}, the first \aclu{SNIPS} algorithm for synchronizing storage peers.
A peer constructs a proof for its stored chunks and sends it to verifier peers.
A verifier queries the proof to identify and subsequently requests missing chunks.
The proof is succinct,~supports membership queries, and requires only a few bits per chunk.

We evaluated our \protocolname algorithm on a cluster of 1000~peers running Ethereum Swarm.
Our results show that \protocolname reduces the amount of synchronization data by three orders of magnitude compared to the state-of-the-art.
Additionally, creating and verifying a proof is linear with the number of chunks and typically requires only tens of µs per chunk.
These qualities are vital for our use case, as we envision running \protocolname frequently to maintain sufficient redundancy consistently.

\end{abstract}


\section{Introduction}
\label{sec:introduction}

\begin{figure}[h]
	\centering
    \includegraphics[width=\columnwidth]{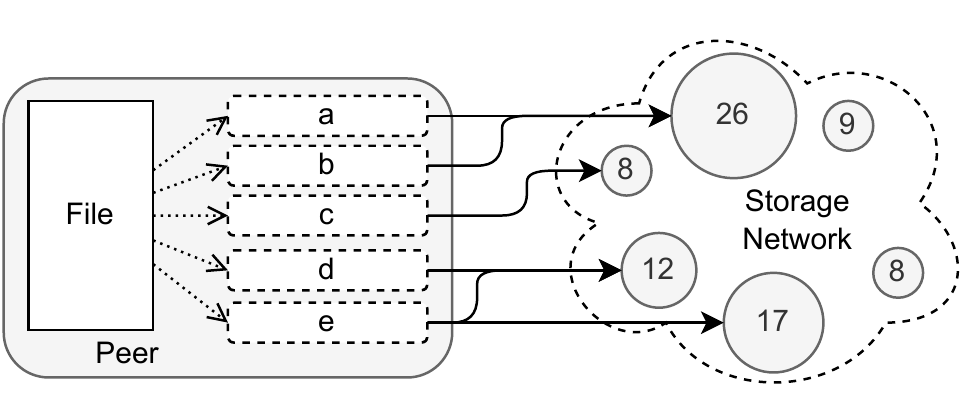}
    \caption{A file is split into chunks and distributed to neighborhoods of different sizes (number in circles).}
	\label{fig:upload}
\end{figure}

\noindent
Decentralized storage systems such as Ethereum Swarm~\cite{bookofswarm}, Filecoin~\cite{filecoin}, and InterPlanetary File System~(IPFS)~\cite{benet2014ipfs} are built on top of large peer-to-peer networks.
The network's peers collaborate in pooling their storage capacity to provide a single storage system.
The peers in Filecoin and Swarm are rewarded to incentivize their contribution of storage and bandwidth to the network.
Peers do not trust each other; instead, they rely on cryptographic primitives to verify the integrity of the data they are tasked to store.
The integrity verification is performed on fixed-size chunks obtained by splitting files.
Chunks are immutable and are identified by their content address, a cryptographic hash of the chunk's content.
\Cref{fig:upload} illustrates the process of uploading a file to a decentralized storage system.

During upload, each chunk is distributed to a neighborhood based on the chunk's content address.
A neighborhood is a cluster of peers with similar addresses.
These neighboring peers are responsible for storing the chunks whose content address is similar to the peers' addresses.
To download a file, peers will similarly attempt to locate the neighborhoods storing the file's chunks.
As \Cref{fig:upload} shows, the neighborhoods can be of varying cardinality.
Thus, the size of a chunk's neighborhood determines the chunk's replication factor.
And since reconstructing a file requires access to all of its chunks, having sufficient redundancy becomes paramount.

Recent studies~\cite{sup,henningsen2020mapping,trautwein2022design,henningsen2020crawling, daniel2022passively} of the network dynamics of Ethereum Swarm and IPFS show that these systems may have tens of thousands of peers connected simultaneously.
However, the studies also show high churn rates, where only a small minority of peers will remain online continuously for 24~hours.
Consequently, peer neighborhoods are constantly changing, potentially compromising the availability of stored files.
One study shows that data loss is inevitable, as it might happen as quickly as within a few days~\cite{sup}.
Hence, storage systems must maintain sufficient redundancy to reduce the likelihood of data loss.
And maintaining redundancy requires \emph{efficient data synchronization} methods.

State-of-the-art protocols for data synchronization in current systems are highly inefficient.
In Swarm, the protocol is called \pullsync, and it relies on peers exchanging long lists of the chunk identifiers they are storing so that the other peer can request the missing chunks.
In an attempt to optimize \pullsync, each peer keeps track of the last synchronization state of its neighbors.
However, as we show in~\Cref{sec:pullsync-inconsistencies}, this optimization is flawed, as inconsistencies can occur.
Filecoin~\cite{filecoin} and IPFS uses a protocol called Bitswap~\cite{bitswapspec} for data synchronization.
Like \pullsync, Bitswap also exchanges long lists of chunk identifiers for content discovery and synchronization.
Bitswap is under active research~\cite{de2021accelerating,beyondbitswap,bitswapimprove}, and various optimizations have been proposed, e.g., GraphSync~\cite{graphsync} and CAR~mirror~\cite{carmirror}.
However, these proposals still rely on exchanging long lists of chunk identifiers.

Previous work~\cite{agarwal2006approximating,byers2004informed,eppstein2011s} on data synchronization in peer-to-peer networks proposed protocols using Bloom filters~\cite{bloom1970space} for approximate reconciliation.
As the name suggests, approximate reconciliation does not guarantee that all chunks are synchronized.
A subset of a file's chunks could be enough to reconstruct the file if it was erasure-coded~\cite{rabin1989efficient}.
However, adding erasure coding to storage systems based on Merkle trees is non-trivial, because of hierarchical dependencies between the content and the metadata~\cite{snarl}.
Hence, we focus on the case where all chunks must be synchronized.

A space-efficient way for peers to ensure they are fully synchronized is to use \ac{PoS} algorithms~\cite{ateniese2007provable}.
\ac{PoS} algorithms allow peers to convince each other what data they are storing without transferring the data itself.
These algorithms typically have three distinct actors; the \emph{challenger}, which issues \ac{PoS} queries; the \emph{prover}, which responds to the queries by creating a proof; and the \emph{verifier}, which verifies the proof.
However, verification typically either completely fails or passes; there is no partial verification.
Even though a peer might have a subset of chunks, the verification will fail as the peer does not have all the chunks.
Moreover, to synchronize data, the verifier would not only need to partially verify proofs but also determine which subset of chunks were missing.
A naive \ac{PoS}-based partial verification approach that creates a proof for each chunk would require communication overhead comparable to exchanging the chunk identifiers themselves.

\subsection{Contributions}
\noindent
This paper presents \protocolname, a novel protocol for data synchronization in decentralized storage systems.
\protocolname uses a novel \ac{PoS}-like construction to generate succinct storage proofs.
The proofs are small, typically only a few bits per chunk, and can be generated and verified efficiently.
Our \ac{PoS} construction is non-interactive, allowing verifiers to make use of the storage proof without a preceding \emph{challenge} phase typically used in \ac{PoS} algorithms.
If both prover and verifier have an identical set of chunks, then both will immediately be convinced of each others state.
Moreover, our construction also allows the verifier to perform membership queries on the proof to determine which chunks are missing, even when the prover and verifier have disjoint sets of chunks.
The verifier may observe false positives for some local chunks that were not part of the prover's proof.
However, our \protocolname protocol can reconcile these false positives.

Peers using the \protocolname protocol generate storage proofs and exchange them with their neighbors.
Upon receiving a storage proof, the peer can query the proof to identify any chunks that it is missing and request the missing chunks from the sender.
False positives are eliminated iteratively with increasingly more accurate proofs.
In summary, a \protocolname proof provides two features;
(1)~it convinces the recipient that the sender has the data, and
(2)~the recipient can query the proof to determine which chunks it is missing.

Our implementation of \protocolname in Ethereum Swarm shows that it is a practical protocol that can be used in decentralized storage systems.
The evaluations show significant bandwidth savings with a reduction in synchronization data transmitted by up to three orders of magnitude compared to current systems.
Moreover, creating and verifying proofs typically requires only tens of microseconds per chunk.
While our evaluation does not compare with Bitswap and its derivatives, \protocolname's approach complements these protocols, and we expect Bitswap would also benefit from our approach.

Our contributions are summarized as follows:
\begin{itemize}
    \item A new \ac{PoS}-like construction for creating storage proofs amenable to partial verification and identification of missing chunks.
    \item The design of \protocolname, a data synchronization protocol for decentralized storage systems, and its implementation in Ethereum Swarm.
    \item A rigorous simulation of bandwidth usage and synchronization overhead and a performance evaluation of \protocolname on a real-world cluster.
\end{itemize}

\section{Preliminaries}
\label{sec:sysmodel}

\noindent
We assume a content-addressed decentralized storage system~\cite{benet2014ipfs,bookofswarm} built on top of a peer-to-peer network, where each \emph{storage peer} is connected to a subset of the peers in the network.
That is, the peers are clustered into small neighborhoods, which collectively share the responsibility of persisting chunks from a specific section of the address space.

We assume no trust in the storage peers and the peer-to-peer network may experience significant churn.
We rely on the storage system's underlying network to route messages between peers.

Each storage peer has access to cryptographic key pairs for signing and verifying digital signatures and we assume that cryptographic primitives cannot easily be circumvented.
Let $\langle m \rangle_{\peershort}$ denote a message $m$ signed by peer $\peershort$.

A peer's public key is further used to generate a unique address for use in the overlay network.
This address serves as the baseline for the connectivity graph for the overlay network, such that peers are more likely to be connected to other peers with similar addresses.

In our algorithm, we assume that all peers have access to a shared source of randomness.
One approach to obtain shared randomness is to use the block hash of a future block in a public blockchain~\cite{futureblock}.
Alternative methods include multi-party computation~\cite{canetti1996adaptively}, Shamir's secret sharing~\cite{syta2017scalable}, verifiable delay function~\cite{vdf}, or verifiable random functions~\cite{vrf}.

We define the following metrics for evaluating \protocolname.

\begin{definition}[Similarity]
    The similarity between two sets, \textit{A} and \textit{B}, is the overlap coefficient~\cite{vijaymeena2016survey}; $\vert A \cap B \vert / \min(\vert A \vert, \vert B \vert)$.
\end{definition}

\begin{definition}[Proof Accuracy]
	Let $M_{i}$ be the set of missing chunks identified when querying a proof.
	Let $M_{p}$ be the set of chunks a peer is missing from a proof.
    The proof accuracy is then $|M_{i}| / |M_{p}|$.
\end{definition}

\subsection{Proof of Storage Model}

\noindent
In our PoS model, there are only provers and verifiers, also referred to as storage peers.
Our PoS-like construction allows peers to
(i)~efficiently synchronize chunks and
(ii)~verify the integrity of their stored chunks with their neighbors.
We assume that peers have full copies of the chunks they store.

Our construction does not provide storage guarantees to clients, as we do not rely on a challenge phase.
We assume clients upload their data to the storage system.
Moreover, chunks in decentralized storage systems are stored on different peers independently of their location within a file.
Thus, our PoS scheme does not bind chunks to a specific location within a file, as it is not required for our use case.
Other protocols can provide these guarantees to clients~\cite{sup,ateniese2007provable}.

\subsection{Threat Model}
\label{sec:threat-model}

\noindent
We consider the threats that can be waged against \protocolname by insider attackers (peers).
We do not consider outsider attackers without credentials, as the storage peers will reject messages from such attackers.
Colluding peers is also not considered; other mechanisms are needed to mitigate collusion attacks (see \Cref{sec:related-works-pos}).
Denial of service attacks is out of scope for this paper, as the underlying storage system should mitigate such attacks.
We focus on forms of attack that aims to compromise the integrity of the storage peers' stored data:

\emph{Replay attacks.}
We consider peers attempting to replay protocol messages to trigger illicit behavior.

\emph{Upload attack.}
After receiving a request for a missing chunk, an attacker may attempt to upload a different chunk.

\emph{Pollution attack.}
Malicious peers may attempt to pollute the storage system by creating and distributing invalid chunks.

\emph{Non-repudiation attack.}
A \protocolname proof can be viewed as a commitment to synchronize the containing chunks.
We consider a non-repudiation attack, where the attacker attempts to modify the containing chunks after distributing the proof.

Moreover, an attacker may attempt to construct a storage proof that results in a false consistency (see~\Cref{sec:false-consistency}).
\Cref{sec:security-analysis} discusses these threats and their mitigations.

\subsection{Swarm}
\label{sec:swarm}

\noindent
Swarm~\cite{bookofswarm} is a global decentralized storage and communication system that distributes stored data to a network of peers.
Swarm's p2p overlay network is based on Kademlia~\cite{maymounkov2002kademlia}, and has more than 2000 active peers~\cite{swarmscan}.

Each peer in Swarm deploys a smart contract, called checkbook, to an EVM-compatible blockchain, e.g., Ethereum or Gnosis.
The contract is used to reward peers with BZZ tokens when they contribute resources to the network.
Specifically, peers pay to download chunks and are rewarded for delivering chunks and forwarding messages.
After deploying the contract, each peer generates a unique peer address used to identify it in the network.
The address is generated by hashing the concatenation of the peer's Ethereum public key, the network identifier, and the hash of the block immediately following the one that deployed the peer's checkbook contract.
The peer is then placed in the neighborhood that shares the longest prefix with the peer's address.

\subsubsection{Data Storage in Swarm}
\label{sec:swarm-data-storage}

Swarm splits files into 4~KB chunks.
A unique \emph{chunk identifier} is derived for each chunk from a cryptographic hash of its content, also referred to as its content address.
Swarm creates a 128-ary Merkle tree where each of the file's chunks is a leaf.
The internal nodes and root of the Merkle tree are also stored in 4~KB chunks and contain a concatenation of the chunk identifiers of their children.

\begin{figure}[h!]
	\centering
	\includegraphics[width=0.8\columnwidth]{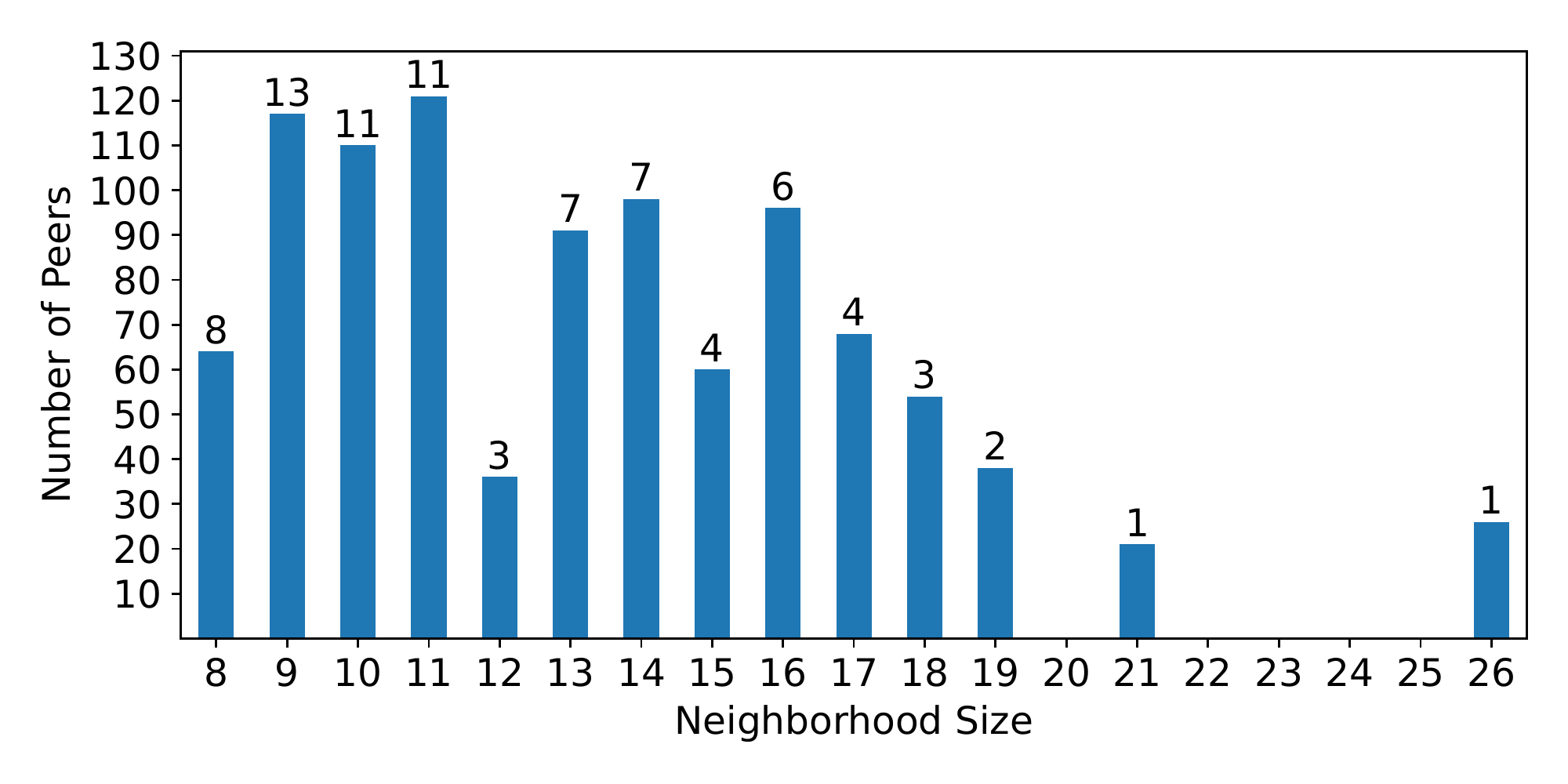}
	\caption{Distribution of neighborhood sizes in our cluster.}
	\label{fig:swarm-network-topology-histo}
\end{figure}

Chunk identifiers and peer addresses share the same address space.
When a chunk is uploaded to the network it is sent to the closest peer, based on the chunk's identifier~\cite{swarmwhitepaper}.
The chunk is then replicated by each peer in the closest peer's neighborhood.
\Cref{fig:swarm-network-topology-histo} shows the distribution of neighborhood sizes obtained in our cluster of 1000~peers.
We observed 81 distinct neighborhoods of various sizes ($n$) with $n \in [8, 26]$. 
As chunks are replicated by all peers in a neighborhood, the varying sizes of the neighborhoods cause the chunks' replication factor to vary accordingly.

\subsubsection{Swarm's \pullsync Protocol}
\label{sec:pullsync}

\begin{figure}[h!]
	\centering
	\includegraphics[width=1\columnwidth]{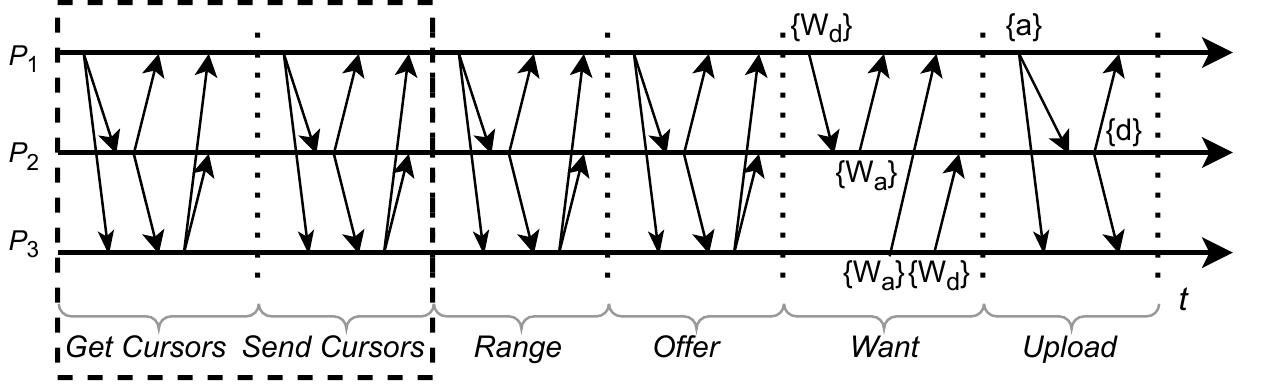}
	\caption{Overview of Swarm's \pullsync protocol.}
	\label{fig:pullsync-overview}
\end{figure}

The goal of Swarm's \pullsync protocol~\cite{bookofswarm,swarmwhitepaper} is to synchronize chunks between neighboring peers.
\pullsync aims to provide eventual consistency among peers within a neighborhood despite churn.
However, \pullsync does not scrub or verify the integrity of the stored data.
\Cref{fig:pullsync-overview} gives an overview of the protocol; the following is a simplified description of the steps.

All peers within a neighborhood run \pullsync with each other.
Each peer monitors the network for topology changes, such as peers joining or leaving, and triggers the \pullsync protocol on such events.
Consider two peers $P_1$ and $P_2$ responding to a topology change.

Peer $P_1$ sends a request to its neighboring peers to get the number of chunks (\emph{cursor}) stored in its neighborhood.
Based on the received cursors, $P_1$ determines if it is missing any chunks, and which peer has the chunks that it may need.

Upon receiving a request for a range of chunks, peer $P_2$ creates an \emph{offer} request containing an array of chunk identifiers in the range.
When $P_1$ receives the offer request, it populates a bit vector with 1s for the chunk identifiers it is missing, and 0s otherwise.
The bit vector is sent back to $P_2$, which will retrieve the missing chunks and deliver them to $P_1$, concluding the synchronization process.

\subsubsection{Inconsistencies in \pullsync}
\label{sec:pullsync-inconsistencies}
\pullsync's inconsistencies can be explained by examining the first three phases in~\Cref{fig:pullsync-overview}.
The \emph{Get Cursors} and \emph{Send Cursors} are only sent the first time two peers synchronize.
After this, each of the peers will keep the other's cursor in memory, and use those for the \emph{Range} phase.

The issue with this arises when a peer deletes a chunk.
Deletion of chunks may be caused by malicious behavior or errors in the \swarmbee client.
However, deletions may also be completely benign, such as due to garbage collection or mechanisms in the incentive layer~\cite{swarmgarbage,bookofswarm}.
In addition, the \swarmbee client exposes an API for the user to delete chunks.

When a peer deletes a chunk, unless it is the last chunk added to the peer, the internal cursors will not be updated.
Therefore, no matter how many times the peer runs the \pullsync protocol, it will not synchronize the deleted chunk.
This causes inconsistencies in which chunks are stored by the neighborhood, and may ultimately result in chunks being lost.

\subsection{\acl{MPHF}}
\label{sec:mpf}

\begin{figure}[h]
	\centering
	\includegraphics[width=0.35\columnwidth]{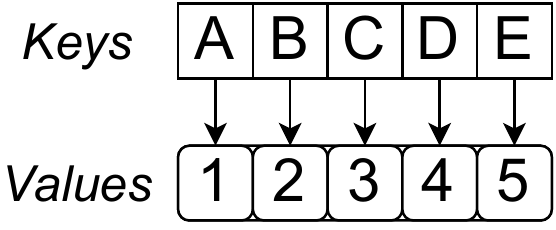}
	\caption{\acl{MPHF} bijective mapping.}
	\label{fig:mphf}
\end{figure}

\noindent
Our enhanced \ac{PoS} construction described in \Cref{sec:pos-construction} uses a \mphf~\cite{esposito2020recsplit,limasset2017fast} to generate storage proofs.
An \mphf is a bijective map from a set of $N$ elements (keys) to the integers~$[1, N]$ (index values).
Each key of the set is mapped to exactly one value, and each value is paired with exactly one key.
\Cref{fig:mphf} illustrates the mapping.

The \mphf allows constructing a space-efficient, collision-free mapping of huge datasets with up to $10^{12}$ keys.
While the theoretical limit is as low as $1.44$ bits per key~\cite{belazzougui2009hash}, practical implementations typically require a few bits per key.
The \mphf is intended for static datasets and has no operations for adding, updating, or deleting elements.
That is, the set of keys must be static and known in advance.
The expected lookup time is $O(1)$, and a mapping can be constructed in $O(N)$ time.

Querying the map for a key that was part of the original set returns the corresponding value in the range $v \in [1, N]$.
However, if the key was not part of the original set, it may return any value in the range $v \in [0, N]$.
When the value $0$ is returned, the key was not part of the original set; however, any other value $v \in [1, N]$ may be a false-positive.
We define a \findchunkproof function for \mphf:

\begin{equation*}
	\label{eq:findproof}
	\findchunkproof(\mathit{elm}):
	\begin{cases}
		0 & \mathit{elm}\ \textrm{definitely not in the set} \\
		v \in [1, N] & \mathit{elm}\ \textrm{might be in the set}
	\end{cases}
\end{equation*}

\section{Enhanced Proof of Storage Construction}
\label{sec:pos-construction}

\noindent
We are now ready to introduce our enhanced non-interactive \ac{PoS} construction.
The construction generates proofs with the following properties:
(1)~it convinces the recipient that the sender has the data, and
(2)~the recipient can query the proof to learn which elements are missing.
The construction consists of two algorithms executed by a \emph{prover} and a \emph{verifier} peer.

\subsection{Prover Algorithm}
\label{sec:prover-algo}

\noindent
The prover uses the \createproof function in \Cref{alg:prover} to generate a \emph{storage proof} for the set of chunks, $\chunkstoreprover$, that the prover stores.
The storage proof is constructed in two steps.

\subsubsection{Create a proof of possession for each stored chunk}
The prover computes a proof of possession for each chunk $a \in \chunkstoreprover$ using \Cref{eq:chunkproof}, as shown in Lines~\ref{alg:prove-snips:chunkloop}-\ref{alg:prove-snips:chunkloopend} of \Cref{alg:prover}.
We refer to this proof as the \emph{chunk proof}~\cite{sup}.
\begin{equation}
	\label{eq:chunkproof}
	\mathrm{ChunkProof}_{a}:\quad cp_{a} = \hash(\mathit{nonce}\ ||\ a)
\end{equation}
The chunk proof is the cryptographic hash of a \emph{nonce} (number used once) concatenated with the chunk's data.
As long as the nonce is unpredictable, the prover cannot claim to store a chunk that it does not, since it must have both the chunk and nonce when generating the proof.
We, therefore, derive the nonce from a shared source of randomness, as discussed in \Cref{sec:sysmodel}.
To that end, the nonce captures the recency of the chunk proof.
We discuss the frequency of proof generation in \Cref{sec:frequency}.
As an optimization, the prover may generate a storage proof for a limited range $[ \varstart, \varend ]$.
In our evaluation, however, we use the entire range of chunks.

\subsubsection{Compress possession proofs into a single storage proof}
The prover compresses the chunk proofs to a succinct storage proof.
We accomplish this by constructing an \mphf (Line~\ref{alg:prove-snips:mphf}), such that each chunk proof $\varchunkproof$ is mapped to a unique index $\varval \in [1,N]$, where $N$ is the number of chunks.

Finally, on Lines~\ref{alg:prove-snips:revmap}-\ref{alg:prove-snips:revmapend}, we build a reverse mapping from the \mphf indices to the chunk identifiers.
\findchunkproof always returns a valid mapping here since the \mphf was constructed from the same chunk proofs.
The prover can use this reverse mapping to identify the missing chunks requested by a verifier.
\begin{align*}
	\label{eq:revindex}
	\varmapvalid & : \varval        \mapsto \vardata.\argaddr & \mathrm{Reverse\ map} \\
	\varmphf     & : \varchunkproof \mapsto \varval           &
\end{align*}

\begin{algorithm}[H]
	\algfontsize
	\caption{Prover: Storage Proof Generation}
	\label{alg:prover}

	\begin{algorithmic}[1]
		\State{$\textbf{Local persistent state at prover:}$}
		\State{$\chunkstoreprover$}
		\Comment{Set of chunks stored by prover}
		\State{$\varmapvalid$}
		\Comment{Reverse map: MPHF index to chunk ID}

		\vspace{0.2cm}

		\MyFunc{$\createproof$}{$\varpublick,\varstart,\varend$}
		\State{$\varbaseproof \setto  \{ \ \}$}
		\State{$\varchunkproofs \setto  \langle \ \rangle$}
		\Comment{Temporary map: chunk proof to chunk ID}
		\ForEach{$\vardata \in \chunkstoreprover : \vardata.\argaddr \in [\ \varstart,\ \varend\ ]$}\label{alg:prove-snips:chunkloop}
			\State{$\varchunkproof \setto \hash(\varpublick \ \concat \ \vardata)$}
			\Comment{Chunk proof}
			\State{$\varbaseproof \setto \varbaseproof \ \cup \ \varchunkproof$}
			\State{$\varchunkproofs \leftbracket \varchunkproof \rightbracket \setto \vardata.\argaddr$}\label{alg:prove-snips:chunkloopend}
		\EndFor

		\State{$\varmphf \setto \new \ \funmphf(\varbaseproof)$}\label{alg:prove-snips:mphf}
		\ForEach{$\varchunkproof \in \varbaseproof$}\label{alg:prove-snips:revmap}
			\Comment{Fill reverse map}
			\State{$\varval \setto \varmphf.\findchunkproof(\varchunkproof)$}
			\Comment{Index of \varchunkproof in proof}
			\State{$\varmapvalid\leftbracket \varpublick \rightbracket \leftbracket \varval \rightbracket \setto \varchunkproofs \leftbracket \varchunkproof \rightbracket$}
			\Comment{Save $\varchunkproof$'s chunk ID}\label{alg:prove-snips:revmapend}
		\EndFor
		\State{$\return\ \leftbracket \ \varmphf, \ \varpublick, \ \varstart, \ \varend \  \rightbracket$}
		\Comment{Storage Proof}
		\EMyFunc
	\end{algorithmic}
\end{algorithm}

\subsection{Verifier Algorithm}
\label{sec:verifier-algo}

\noindent
Next, we explain how the verifier can use a storage proof from the prover to identify chunks that the verifier may be missing using the \findmissing function shown in \Cref{alg:verifier}.

Given a storage \argproof, the verifier computes chunk proofs for each local $\vardata \in \chunkstoreverifier$ over the proof's range.
These chunk proofs are then used to query the prover-generated \mphf using the \findchunkproof function (Line~\ref{alg:maintain-snips:findchunk}).
Thus, if the chunk proof is in the \mphf proof, \findchunkproof returns an $\varval \in [1,N]$.
Otherwise, $\varval = 0$ is returned.
We use $\varval$ as an index into the \varexpectedvals array.
On Line~\ref{alg:maintain-snips:foundchunk}, the verifier decrements the \mbox{$\varval$-th} entry in \varexpectedvals to indicate that the local $\vardata \in \argproof$.
Once the verifier has processed all the chunk proofs, we use \varexpectedvals to identify the missing chunks and collisions as follows:

\vspace{-.3cm}
\begin{equation*}
	\label{eq:foundorcollision}
	\varexpectedvals[\varval]
	\begin{cases}
		=  1 &  \textrm{verifier \textbf{m}issing chunk for}\ \varval \\
		=  0 &  \textrm{verifier \textbf{f}ound chunk locally} \\
		<  0 &  \textrm{verifier has \textbf{c}ollision(s) for}\ \varval
	\end{cases}
\end{equation*}

That is, entries in \varexpectedvals whose value is still equal to~1 must be missing (Line~\ref{alg:maintain-snips:missingchunk}), and we add them to the \varmissing set.
Entries whose value equals~0 are not missing and require no further processing.
For entries with a negative value (Line~\ref{alg:maintain-snips:collision}), however, the verifier found two or more chunk proofs that mapped to the same index value, i.e., a collision.
We discuss this issue in \Cref{sec:collision}.

It is essential to recognize that the verifier cannot specify the missing chunks in the form of chunk identifiers or chunk proofs.
Instead, the \varmissing set contains index values that the prover can use to look up the missing chunks' identifiers in the \varmapvalid reverse mapping.
Recall that the prover built the \varmapvalid on Lines~\ref{alg:prove-snips:revmap}-\ref{alg:prove-snips:revmapend} of \Cref{alg:prover}.

\begin{algorithm}[H]
	\algfontsize
	\caption{Verifier: Find Missing Chunks}
	\label{alg:verifier}

	\begin{algorithmic}[1]
		\State{$\textbf{Local persistent state at verifier:}$}
		\State{$\chunkstoreverifier$}
		\Comment{Set of chunks stored by verifier}

		\vspace{0.2cm}

		\MyFunc{FindMissingChunks}{$\argproof$}
		\State{$\varproofsize \setto |\ \argproof.\varmphf\ |$}
		\Comment{Number of chunks in proof}
		\State{$\varexpectedvals \setto \leftbracket 1,\ldots,1 \rightbracket^{\varproofsize}$}
		\Comment{Initially, all chunks are missing}

		\vspace{0.1cm}

		\ForEach{$\vardata \in \chunkstoreverifier : \vardata.\argaddr \in [\ \argproof.\varstart,\ \argproof.\varend\ ]$}
			\State{$\varchunkproof \setto \hash(\argproof.\varpublick \ \concat \ \vardata)$}
			\State{$\varval \setto \argproof.\varmphf.\findchunkproof(\varchunkproof)$}
			\label{alg:maintain-snips:findchunk}
			\Comment{Index of \varchunkproof in proof}
			\If{$\varval \neq 0$}
				\Comment{Might be in the set}
				\State{$\varexpectedvals[\varval] \setto \varexpectedvals[\varval] - 1$}
				\label{alg:maintain-snips:foundchunk}
				\Comment{Found chunk or collision}
			\EndIf
		\EndFor
		\State{$\varmissing \setto \{ \ \}$}
		\State{$\varnewproof \setto \false$}
		\For{$\varinterval \setto \leftbracket 1,\ \varproofsize \rightbracket$}
			\If{$\varexpectedvals[\varinterval] = 1$}
				\State{$\varmissing \setto \varmissing \cup \varinterval$}
				\label{alg:maintain-snips:missingchunk}
				\Comment{Missing chunk}
			\ElsIf{$\varexpectedvals[\varinterval] < 0$}
				\State{$\varnewproof \setto \true$}
				\label{alg:maintain-snips:collision}
				\Comment{Found collision}
			\EndIf
		\EndFor
		\State{$\return\ \leftbracket \ \varmissing, \ \varnewproof \ \rightbracket$}
		\Comment{Missing chunks}

		\EMyFunc
	\end{algorithmic}
\end{algorithm}

\subsection{Collisions}
\label{sec:collision}

\noindent
This section explains why collisions can occur when verifying a proof.
Recall that \mphf's \findchunkproof function may return a false positive, which can happen when a chunk not part of the original set responds with ``\emph{might be in the set}.''
Formally:
\begin{definition}[False Positive]
    Let $\chunkstoreprover$ be the set of chunks used to construct the storage \argproof and let $\varchunkproof \notin \chunkstoreprover$.
    Then we have a false positive if $\findchunkproof(\varchunkproof) > 0$.
\end{definition}

Given that \mphf admits false positives, it is also possible to have a collision.
A collision happens when multiple chunks map to the same index value in the proof.
Having a collision means that there is at least one false positive.
Formally:
\begin{definition}[Collision]
    Let $\chunkstoreprover$ be the set of chunks used to construct the storage \argproof,
    and let $\varchunkproof_a \in \chunkstoreprover$ and $\varchunkproof_b \notin \chunkstoreprover$
    such that $\varchunkproof_a \neq \varchunkproof_b$.
    Then we have a collision if:
    \[
        \findchunkproof(\varchunkproof_a) = \findchunkproof(\varchunkproof_b)
    \]
\end{definition}

The prover and verifier independently build their chunk proofs from $\chunkstoreprover$ and $\chunkstoreverifier$, respectively.
Since $\chunkstoreverifier$ may differ from $\chunkstoreprover$, and the verifier use chunk proofs from $\chunkstoreverifier$ to query the prover's storage \argproof, the verifier may observe a collision if two or more chunk proofs map to the same index value (Line~\ref{alg:maintain-snips:findchunk} in \Cref{alg:verifier}).

\begin{figure}[H]
    \centering
	\includegraphics[width=0.8\columnwidth]{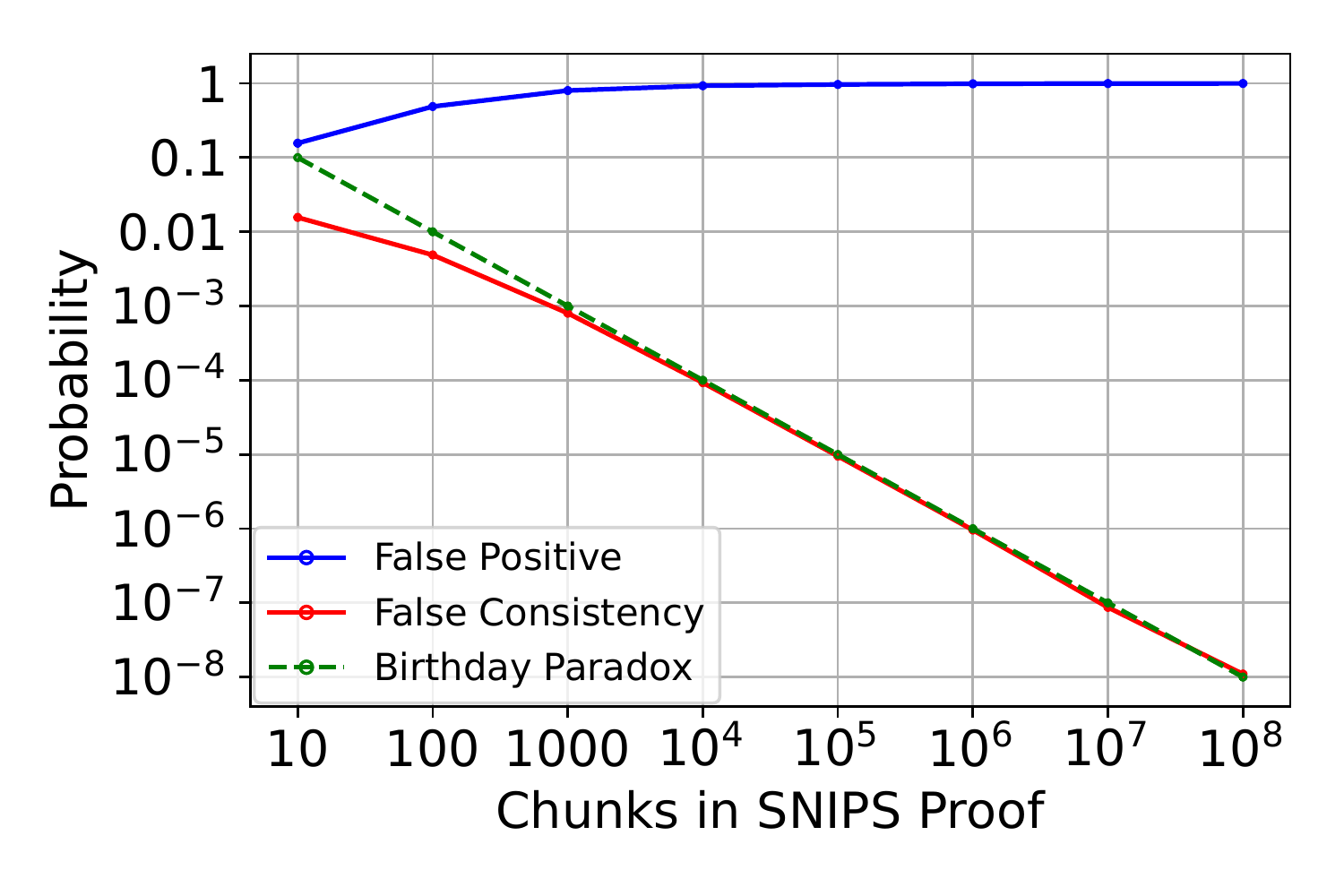}
	\caption{Probability of false positive and false consistency as a function of the number of chunks in the proof. The dotted green line shows an estimation of false consistency.}
	\label{fig:eval-falsepositive}
\end{figure}

We simulated a single peer verifying proofs of different sizes to determine the probability of false positives.
As we can see from~\Cref{fig:eval-falsepositive}, the probability of observing false positives increases with the number of chunks in the proof.
Given a proof of 10\textsuperscript{8} chunks, the probability of observing a false positive is 99.78~\%.
Although false positives and collisions can occur, their impact is limited as \protocolname uses a strategy of repeated proof generation until all missing chunks are discovered.
We describe the strategy in~\Cref{sec:protocol} and evaluate it in~\Cref{sec:eval-similarity}.
In fact, we conjecture that even if the proof had a 100~\% probability of false positives, \protocolname would be able to discover all missing chunks.
We reason by observing that even if false positives exist, it is still possible to discover missing chunks --- unless these false positives result in false consistency, which we discuss next.

\subsection{False Consistency}
\label{sec:false-consistency}
\noindent
As we explained in \Cref{sec:collision}, it is possible that the verifier peer can observe a collision if two or more chunk proofs map to the same index value.
While such collisions are unlikely, they are easy to detect and recover from.
In this section, we discuss \emph{false consistency}---a related problem that may only occur when similarity is below 1, i.e., both peers have chunks that the other does not.

\begin{definition}[False Consistency]
    When a peer believes it has no missing chunks after querying a proof.
    However, at least one chunk used to query the proof gave a false positive, and the value of the false positive exactly matches the last remaining expected value.
\end{definition}

To explain false consistency, consider a proof exchange between two peers, $P_1$ and $P_2$.
Peer $P_1$ stores chunks $\{a, b, c\}$ and peer $P_2$ stores chunks $\{a, b, d\}$, where chunk $c \neq d$.
When peers $P_1$ and $P_2$ evaluate each other's \argproof to identify missing chunks, it is possible that they both have a false positive, resulting in two distinct mappings, as illustrated here:
\begin{gather*}
    P_1:  [a \mapsto 3, b \mapsto 1, d \mapsto 2]\\
    P_2:  [a \mapsto 1, b \mapsto 2, c \mapsto 3]
\end{gather*}

In this case, there are no collision mappings nor any mapping to $0$.
That is, $P_1$ believes it has all chunks in the storage \argproof sent by peer $P_2$ and vice versa.
Thus, both peers falsely conclude that they have the same set of chunks.

\Cref{fig:eval-falsepositive} shows the results of simulating a single peer verifying proofs of different sizes to determine the probability of false consistency.
The probability of false consistency decreases linearly with an increasing number of chunks in the proof, and becomes negligible for a large number of chunks.
As an example, for a proof with $1000$ chunks, corresponding to a peer storing only 4~MB, the probability of observing a false consistency is less than 0.09~\%.

Fortunately, as the number of chunks increases, observing a false consistency is unlikely in practice.
By using a different nonce for each proof, there is a very high probability that the false consistency will be resolved in the next proof exchange.

We can model the probability of a false consistency using the \emph{birthday paradox}~\cite{good1950probability}, shown as a green line in~\Cref{fig:eval-falsepositive}.
The birthday paradox is a classic technique used to find the probability of a collision of two or more randomly chosen elements in a set.
In our case, we let $N$ be the cardinality of the set of all possible index values in the proof, and let $k$ represent the number of peers exchanging proofs.
The equation is given in~\Cref{eq:birthday-problem} as follows:
\begin{equation}
    \label{eq:birthday-problem}
    P({c}) = 1 - \frac{N!}{(N - k)!} \frac{1}{N^k}
\end{equation}

\noindent
As there are only 2 peers exchanging proofs, we can simplify to get the expression $P(c)|_{k=2} = 1 / N$.

\section{The \protocolname Protocol}
\label{sec:protocol}

\noindent
This section presents \protocolname, a data synchronization protocol for decentralized storage systems.
\protocolname allows storage peers to perform periodic checks for missing chunks with negligible bandwidth overhead and without a preceding challenge phase.

\protocolname provides a mechanism for peers to exchange storage proofs, and to iteratively improve synchronization accuracy.
Used together with the \ac{PoS} construction in \Cref{sec:pos-construction}, we can use fewer bits per chunk in the storage proof.
Thus, \protocolname helps to strike a balance between the size of the storage proof versus the iterations needed to synchronize the peers.

\begin{figure}[h!]
	\centering
	\includegraphics[width=0.8\columnwidth]{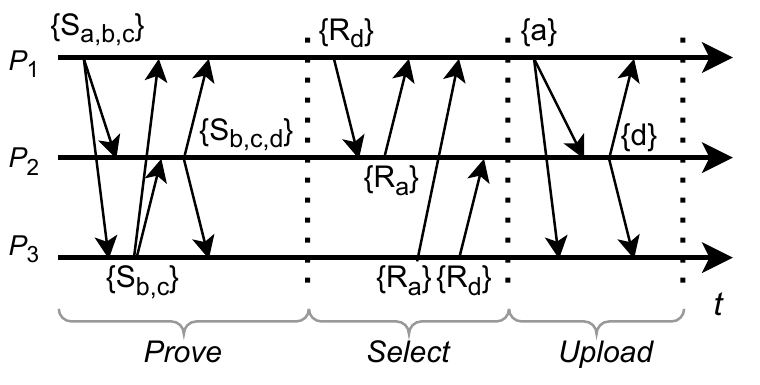}
	\caption{Space-time diagram of \protocolname's phases.}
	\label{fig:snips-overview}
\end{figure}

We first give an overview of \protocolname' three phases, as shown in \Cref{fig:snips-overview}.
While \Cref{fig:snips-overview} illustrates the different phases as if they are synchronized, \protocolname operates entirely asynchronously.
Each peer can progress at their own pace.

In the \emph{Prove} phase, each peer constructs a storage proof for its chunks.
Peer $P_1$ constructs the proof $S_{a,b,c}$, representing that it stores the chunks $\{a,b,c\}$.
Similarly, $P_2$ constructs $S_{b,c,d}$, and $P_3$ constructs $S_{b,c}$.
The peers $P_1$, $P_2$, and $P_3$ send their storage proofs to the other two.

Upon receiving the storage proof $S_{b,c,d}$ from $P_2$, $P_1$ detects that it is missing chunk $\{d\}$ and requests it from $P_2$ with the $R_{d}$ message in the \emph{Select} phase.
Similarly, $P_2$ detects that it is missing chunk $\{a\}$ and requests it from $P_1$.
In the same way, $P_3$ detects that it is missing chunks $\{a,d\}$ and requests them from $P_1$ and $P_2$, respectively.
Finally, the requested chunks are uploaded in the \emph{Upload} phase, and the storage redundancy is recovered.
Next, we explain each phase in more detail.

\subsection{Prove Phase: Storage Proof Generation}

\noindent
In the Prove phase, shown in \Cref{alg:prove-snips}, each peer constructs a storage proof for the set of chunks that the peer stores using the \createproof function in \Cref{alg:prover}.
The proof is then digitally signed to preserve its integrity and to link it to the prover's public key.

The Prove phase is triggered via the $\msgnewproof$ message in two ways:
periodically from the local peer (see \Cref{sec:frequency}) or
from another peer to quickly recover from a collision (see Line~\ref{alg:maintain-snips:requestnewproof} in \Cref{alg:maintain-snips}).

\begin{algorithm}[H]
	\algfontsize
	\caption{Prove Phase: Storage Proof Generation}
	\label{alg:prove-snips}

	\begin{algorithmic}[1]
		\Receive{$\langle \ \msgnewproof \ | \ \varpublick, \varstart, \varend \ \rangle$}
		\State{$\argproof \setto \createproof(\varpublick, \varstart, \varend)$}
		\State{$\send \ \langle \ \msgprove \ | \ \argproof\ \rangle_{\peershort}$}
		\Comment{Signed by peer $\peershort$}
		\EndReceive
	\end{algorithmic}
\end{algorithm}

\subsection{Select Phase: Request Missing Chunks}
\label{sec:select}

\noindent
In the Select phase, shown in \Cref{alg:maintain-snips}, a storage peer expects to receive \msgprove messages from other peers periodically.
Upon receiving a storage \argproof, the peer calls the \findmissing function to identify any potentially missing chunks.
If the peer detects missing chunks, it can request them by sending a \msgselect message to the peers that created the corresponding storage proofs.

As we mentioned in \Cref{sec:verifier-algo}, evaluating the storage proof can result in collisions.
The \varnewproof flag indicates this.
Thus, to reconcile collisions, we trigger a new Prove phase (Line~\ref{alg:maintain-snips:requestnewproof}).
However, we wait until the peer's requested chunks have been received, as indicated by the \msguploaddone message.
Depending on the peer neighborhood's storage size to synchronize and the similarity between prover and verifier, it may require multiple iterations to reconcile all collisions.

Since the verifier only knows the prover's missing chunk indexes, we may request the same chunk from multiple peers.
Hence, we sequentially process the proofs from different peers to avoid duplicate chunk uploads.
However, we request chunks from the same peer concurrently.
Upon receiving chunks, we evaluate them against all proofs.

\begin{algorithm}[H]
	\algfontsize
	\caption{Select Phase: Request Missing Chunks}
	\label{alg:maintain-snips}

	\begin{algorithmic}[1]
		\Receive{$\langle \ \msgprove \ | \ \argproof \ \rangle_{\peershort}$}
		\Comment{Verified proof from peer $\peershort$}

			\State{$\leftbracket \ \varmissing, \ \varnewproof \ \rightbracket \setto \ \findmissing(\argproof) $}
			\If{$\varmissing = \emptyset$}
				\State{$\return$}
				\Comment{Nothing is missing}
			\EndIf

			\State{$\reply \ \langle \ \msgselect \ | \ \argproof.\varpublick,\ \varmissing \ \rangle$}
			\label{alg:maintain-snips:requestmissing}
			\Comment{Request chunks}

			\vspace{0.1cm}

			\If{$\varnewproof$}
			\Receive{$\langle \ \msguploaddone \ \rangle $}
				\State{$\reply \ \langle \ \msgnewproof \ | \ \argproof.\varpublick, \ \argproof.\varstart, \ \argproof.\varend \ \rangle$}
				\label{alg:maintain-snips:requestnewproof}
			\EndReceive
			\EndIf
		\EndReceive
	\end{algorithmic}
\end{algorithm}

\subsection{Upload Phase: Send Missing Chunks}

\noindent
The Upload phase presented in \Cref{alg:upload-snips} is simple.
In this phase, the prover responds to \msgselect messages by uploading the missing chunks to the verifier.

The prover peer that created the storage proof stores the \varmapvalid reverse mapping for the nonce associated with the storage proof.
Hence, to facilitate the upload, the prover uses the \varmapvalid mapping.
We find the chunk identifier in the reverse mapping for each index value in the request's \varmissing set.
Then we use that chunk identifier to find the chunk in the local storage, $\chunkstoreprover$, and send that chunk to the requesting verifier peer.
Finally, the \msguploaddone message signals to the verifier peer that the upload is done.

\begin{algorithm}[H]
	\algfontsize
	\caption{Upload Phase: Send Missing Chunks}
	\label{alg:upload-snips}

	\begin{algorithmic}[1]
		\State{$\textbf{Local persistent state at prover:}$}
		\State{$\chunkstoreprover$}
		\Comment{Set of chunks stored by prover}
		\State{$\varmapvalid$}
		\Comment{Reverse map: MPHF value to chunk ID}

		\vspace{0.2cm}
		\Receive{$\langle \ \msgselect \ | \ \varpublick, \ \varmissing \ \rangle$}
			\ForEach{$\varval \in \varmissing$}
				\State{$\varchunkid \setto \varmapvalid\leftbracket \varpublick \rightbracket \leftbracket \varval \rightbracket$}
				\State{$\vardata \setto \{c : c \in \chunkstoreprover \wedge c.\argaddr = \varchunkid\}$}
				\State{$\reply \ \langle \ \msgupload \ | \ \vardata \ \rangle$}
			\EndFor
			\State{$\reply \ \langle \ \msguploaddone \ \rangle$}
		\EndReceive
	\end{algorithmic}
\end{algorithm}

\subsection{Proof Generation Frequency}
\label{sec:frequency}

\noindent
We briefly explore how to limit the execution frequency of \protocolname.
For example, in a decentralized storage system, we envision \protocolname running approximately once per day.

To accomplish this, we can configure protocol execution to follow the block generation frequency of a public blockchain.
For example, we can trigger the Prove phase by extracting a new nonce from the blockchain every $k$-th block.
The nonce can be the hash of the blockchain's current block header.
Since blocks are produced at regular intervals, we can limit protocol execution to the desired frequency.
For example, Ethereum has an average block time of around 12 seconds~\cite{ethblocktime,ethslottime}.
Thus, to generate a storage proof once a day, we can trigger execution every 7200-th block
($24\mathrm{h}\cdot 60\mathrm{m}\cdot 60\mathrm{s} / 12\mathrm{s}$).

Having the peers issue proofs on the same interval for the same nonce allows the peers to amortize the cost of proof generation.
We leave it for future work to explore how the peers can coordinate the proof generation to save costs.

\subsection{Security Analysis and Mitigation Strategies}
\label{sec:security-analysis}

\noindent
This section provides a cursory security analysis of \protocolname and some mitigations.
We note that the underlying storage system already mitigates several attack vectors.
For example, we do not consider omission and freeloading in this work, since the underlying storage system should mitigate these attacks.
More work is needed to understand if \protocolname exposes additional attack vectors for such attacks and how to mitigate them.
We analyze the storage peer integrity attacks outlined in~\Cref{sec:threat-model} in the following.

\emph{Replay attack.}
An attacker can wage a replay attack by sending the same Prove or Select message multiple times.
A peer may process these replayed messages without affecting the storage peer's integrity.
However, such attacks impact the protocol's performance due to repeated work and added network traffic.
We note that processing NewProof and Prove messages are more costly than processing Select messages (see~\Cref{fig:eval-computation}).
Thus, we may also rely on the storage system's mechanisms for rate-limiting peers.

An alternative mitigation strategy is discarding replayed messages based on the nonce and peer address.
With this strategy, peers cannot retransmit messages for the same nonce and must wait for the next synchronization round instead.

\emph{Upload attack.}
A misbehaving peer may upload different chunks than those requested in the Select phase.
To mitigate this, the receiving peer first checks that the uploaded chunks belong to its neighborhood.
Next, the peer checks that the uploaded chunks map to the expected index values in the storage proof previously received from the uploading peer.

To illustrate, consider the example in \Cref{fig:snips-overview}, where $P_2$ sent a request for $R_a$ and then received chunk $a$ from $P_1$.
Thus, $P_2$ can check that $a$ maps to $R_a$ in the original storage proof $S_{a,b,c}$ sent by $P_1$.
If the mapping is incorrect, then $P_2$ rejects the chunk.
If $P_1$ sent one of the other chunks in $S_{a,b,c}$ that $P_2$ already has, then $P_2$ will also reject the chunk.

\emph{Pollution attack.}
A misbehaving peer may attempt to pollute the storage system by creating invalid chunks, e.g., chunks not uploaded by a client.
The attacking peer may include the invalid chunks in the storage proof and hope that other peers will request them.
\protocolname does not prevent this attack; we leave it to the underlying storage system to mitigate it.
A trivial mitigation strategy requires peers to embed the client's signature with each upload.

\emph{Non-repudiation attack.}
When a peer distributes a storage proof, it commits to the chunks in the proof.
However, an attacker could attempt to deceive other peers into accepting a non-verified chunk by exploiting a collision in the proof sent during the Prove phase.

Consider a scenario with two peers, $P_1$ and $P_2$, where $P_1$ is the attacker.
Let $P_1$ have the chunks $\{a,b\}$, and $P_2$ have $\{a\}$, but not $b$.
At the start of the protocol, $P_1$ creates a storage proof $S_{a,b}$ and sends it to $P_2$.
Next, $P_2$ sends a request $R_b$ for chunk $b$ from $P_1$.
Assume now that $P_1$ can find a trojan chunk $b' \neq b$, such that $\findchunkproof(\varchunkproof_{b'}) = \findchunkproof(\varchunkproof_b)$.
Finally, $P_1$ can upload $b'$, and $P_2$ would accept $b'$ as a valid chunk instead of $b$, even though $b'$ is not contained in $S_{a,b}$.

We argue that this attack is difficult to perform in practice for typically sized storage systems.
First, the attacker must find a candidate chunk $b'$ that collides with $b$.
Second, $b'$ must have the same chunk identifier prefix as $b$; otherwise, $b'$ would belong to a different neighborhood, and $P_2$ would not accept it.
This involves hashing over each candidate $b'$ until the resulting chunk identifier has a shared prefix.
Then, the attacker must compute the chunk proof of $b'$, as described in~\Cref{sec:prover-algo}, and then determine if it collides with $b$.
If the chunk proof does not collide, the attacker moves to the next candidate $b'$.

Given a prefix length of $l$ bits and a proof of $n$ chunks, the probability of finding a chunk $b'$ with both a shared chunk identifier prefix and that collides with $b$ is $1/(2^l n)$.
Hence, given a 16-bit prefix and a proof of 1 million chunks, the probability of finding a trojan chunk is $1.53\cdot 10^{-11}$.

To further mitigate the non-repudiation attack, e.g., for smaller-sized storage systems, we can include a ``proof checksum'' in the storage proof.
The proof checksum is a hash of the chunk proofs in the storage proof, i.e., $\hash(\varchunkproof_1 \ \concat \ \varchunkproof_2 \ \concat \ \dots)$.

We mention briefly that an attacker may also attempt to exploit the possibility of a false consistency.
However, this attack would be at least as costly as the non-repudiation attack.

\section{Implementation}
\label{sec:impl}
\noindent
We implemented \protocolname as a package in \swarmbee \swarmbeeversion.
The implementation comprises about 800 lines of Go code plus about 1900 lines for benchmarking, testing, and metrics collection.
Our implementation is based on the protocol described in~\Cref{sec:protocol}.
For the \mphf, we used the implementation from~\cite{gobbhash} with the recommended expansion factor $\gamma=2$.

In this section, we describe the implementation details of the protocol.
We start by describing the protocol messages and the data structures used by the prover and verifier.
Next, we describe how we implemented the prover and verifier, including some optimizations.
Lastly, we describe some aspects of our evaluation framework.

\subsection{Protocol Messages}
\label{sec:protocol-messages}
\noindent
We defined \protocolname' four messages using protocol buffers~\cite{protobuf}.
First, the Proof message contains
 an 8-byte nonce,
 two 32-byte start and end fields,
 a 4-byte entry for the proof's length, and
 a variable-length entry for the \mphf proof itself.
Additionally, the proof is signed, adding another 97 bytes for a Swarm-specific signature.
The Swarm-specific signature allows peers to verify that the proof's creator is within the same neighborhood.

The Select message only contains an 8-byte nonce and a bit vector.
The bit vector allows us to efficiently select multiple chunks in the same message.

The NewProof message only contains an 8-byte nonce.
This is a deviation from the protocol described in~\Cref{sec:protocol}, where the NewProof message also contains the start and end of the proof.
This is because, in our implementation, the prover keeps track of the start and end, and it is thus unnecessary to include them in the message.
Lastly, the UploadDone message is only used as a signal and does not contain any data.

\subsection{Prover}
\noindent
At some point, each peer in the network will act as a prover.
Therefore, as soon as a peer starts, it will begin listening for new blocks to be added to the blockchain.

The other trigger to initiate proof generation is when a peer receives a NewProof message.
In this case, the peer will check if a proof for the same nonce already exists in the \varmapidproof cache.
If so, the peer will obtain the proof from the \varmapidproof cache instead of generating a new one.
Hence, we can amortize the cost of generating the chunk proofs by caching the chunk proofs.
This is particularly useful if peers issue proofs on the same interval for the same nonce, e.g., every $7200$-th block.

\subsection{Verifier}
\noindent
The verifier uses the same \varmapidproof cache as the prover.
In addition, the verifier uses a bit vector to represent the missing chunks in the Select message instead of a list of indexes.
We observe the benefit of using a bit vector increase as the number of missing chunks increases.
The added overhead of the bit vector makes it comparable to sending the index for only a single missing chunk.

\subsection{Blockchain as a Shared Randomness Source}
\noindent
To obtain a shared randomness source, we implemented a small probe that listens for new blocks from the Ethereum blockchain.
Once a new block is received, the probe will determine if it is the next one in the interval by comparing its block number with the next expected block number.
If the block number is greater or equal to the next expected block number, the probe will query the blockchain for the block hash of the next expected block number.
Thus, even if we were slow to query the blockchain for new blocks, once we catch up, we will use the same block hash as the other peers for the nonce.

\subsection{Evaluation Framework}
\noindent
We give a brief overview of the evaluation framework we implemented to evaluate the performance of \protocolname and \pullsync.
To collect metrics, we use the Prometheus~\cite{prometheus} monitoring system.
We registered interesting metrics for \protocolname and \pullsync, such as the number of messages sent, the number of bytes in messages, the time to process messages, and more.
In addition, as the protocols do not wait for chunks to be synchronized before terminating, we used Prometheus to monitor the peers' activity to determine when synchronization was completed.

In both \protocolname and \pullsync, communication is contained within neighborhoods.
To ensure that we only collected metrics from peers in the neighborhoods and did not have any outside interference, we added an API endpoint to the \swarmbee client that allowed us to query the neighborhood of a peer.

\section{Evaluation}

\noindent
This section presents our evaluation of \protocolname.
Using a real-world Swarm deployment on our cluster, we measured \protocolname's performance in terms of the amount of transmitted data, bandwidth savings, synchronization time, computation time, and per-chunk bandwidth requirements.
We compared our measurements to Swarm's \pullsync protocol.
We evaluate full synchronization runs in scenarios with \emph{chunk loss}~(CL) and \emph{adding new chunks}~(CA).

Our results show that \protocolname uses up to three orders of magnitude less synchronization data than \pullsync.
\protocolname's computational overhead shows that it is a practical protocol requiring only tens of microseconds per chunk to create and verify proofs.
We also simulated the performance of \protocolname in the most challenging synchronization scenarios, where the \emph{similarity} between peers is $0$.
The simulations show that \protocolname performs well under challenging conditions and can always efficiently synchronize peers.
We varied the number of chunks so that the neighborhoods' total storage ranges from $1$~MB to $1000$~MB and up to 10~GB for some experiments.

\subsection{Experimental Setup}

\noindent
The experiments were conducted on a cluster of 30 physical machines running Ubuntu~18.04.4~LTS.
Each machine has 32~GB RAM, an Intel Xeon E-2136 3.30~GHz CPU, a 1.5~TB SSD disk, and 10~Gbit/s NIC.
We used the cluster to run a large network of 1000 Swarm peers.
Using Kubernetes~\cite{kube} and Helm~\cite{helm}, we distributed the Swarm peers on 28 machines, using one to host a private Ethereum network and one for managing the experiment execution.

The distribution of our Swarm network is shown in \Cref{fig:swarm-network-topology-histo}.
There are 81~neighborhoods whose size varies between $8$ and $26$ peers.
Since \protocolname's synchronization operations are confined to a single neighborhood, we evaluated the protocol with neighborhoods of $8, 17$, and $26$ peers.
Hence, a rough Fermi estimate for the system-wide bandwidth savings would be $81\times$ the individual savings of one neighborhood.

\subsection{Consistent \pullsync}
\label{sec:consistent-pullsync}
\noindent
Comparing the performance of \protocolname and \pullsync is challenging due to \pullsync's inconsistencies.
In addition, \protocolname has fewer phases than \pullsync, and the initiator of \protocolname wants to upload chunks, while the initiator of \pullsync wants to download chunks.
To compensate, our duration measurements are taken from when the protocol is initiated until all peers are fully synchronized.
However, this is not without caveats, as \pullsync gossips new chunks to its neighbors, reducing the chunk distribution time while potentially wasting bandwidth due to duplicate chunk uploads.
While \protocolname could trivially implement a similar optimization, our priority is to reduce the amount of transmitted data.

For data transmission measurements, we were able to isolate data synchronization traffic from chunk uploads.
However, to evaluate the chunk loss scenario, we modified \pullsync so that it does not cause inconsistencies.
When new chunks are added, we compare \protocolname with vanilla \pullsync.

\subsection{Real-world Comparison of \protocolname and \pullsync}
\label{sec:eval-realworld}

\noindent
We first compare \protocolname with \pullsync by measuring the \emph{data transmitted} and \emph{time to synchronize} a neighborhood under the same conditions in our cluster.
We repeated the experiments $10$ times for each configuration, and fixed the \emph{similarity} between peers to $1$, meaning that $\vert A \cap B \vert = \min(\vert A \vert, \vert B \vert)$.

\begin{figure*}[ht]
	\centering
	\includegraphics[width=\textwidth]{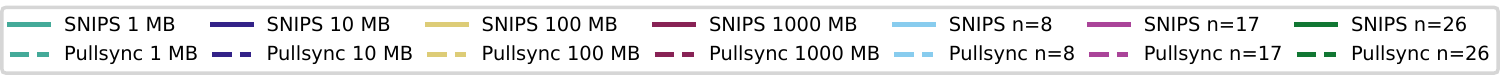}
	\begin{subfigure}[t]{0.245\textwidth}
		\centering
		\includegraphics[width=\columnwidth,trim={1.02cm 1cm 0cm 0}]{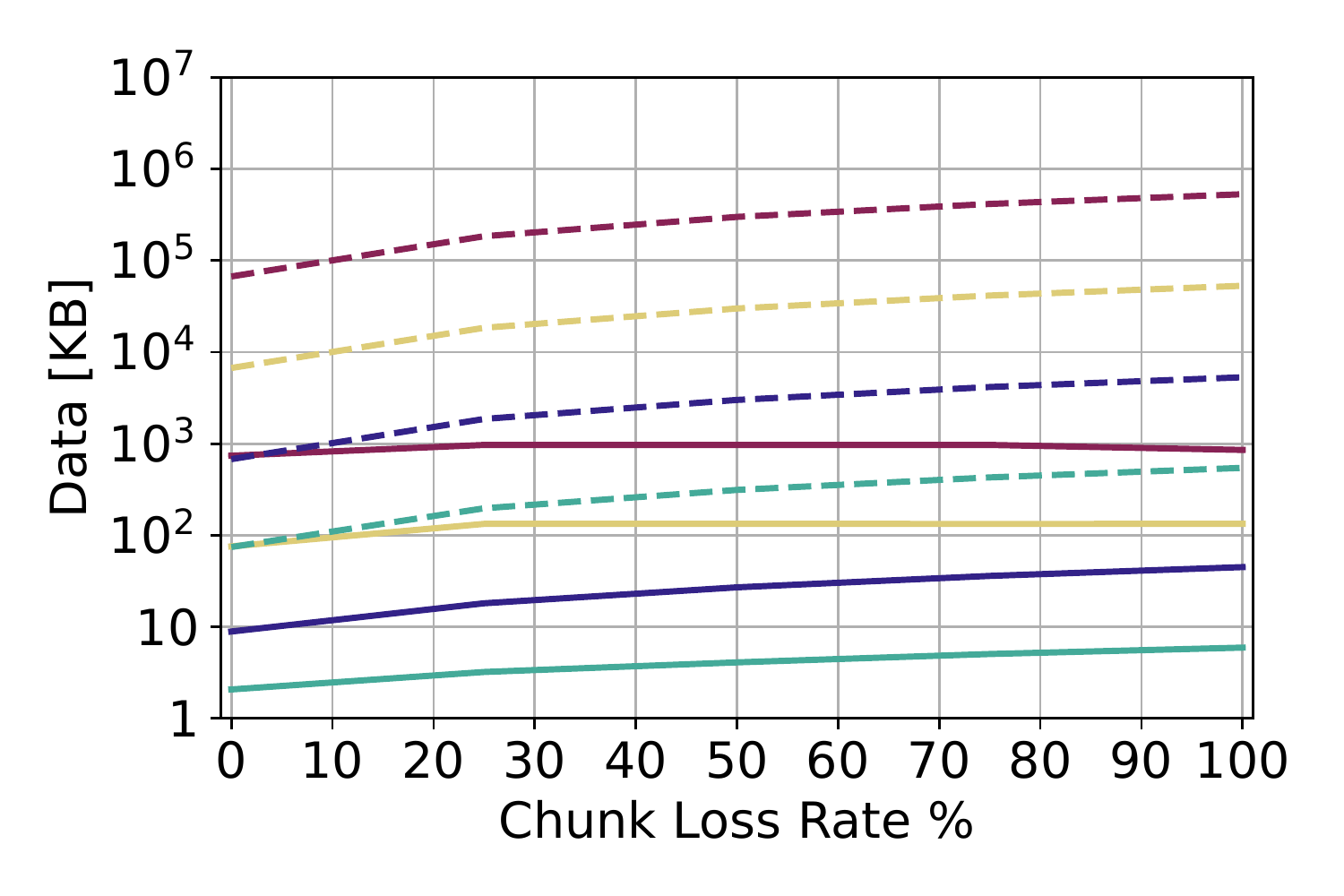}
		\caption{Data transmitted, $n$=8, CL}
		\label{fig:net-bandwidth-cl-97}
	\end{subfigure}
	\begin{subfigure}[t]{0.245\textwidth}
		\centering
		\includegraphics[width=\columnwidth,trim={1.02cm 1cm 0cm 0}]{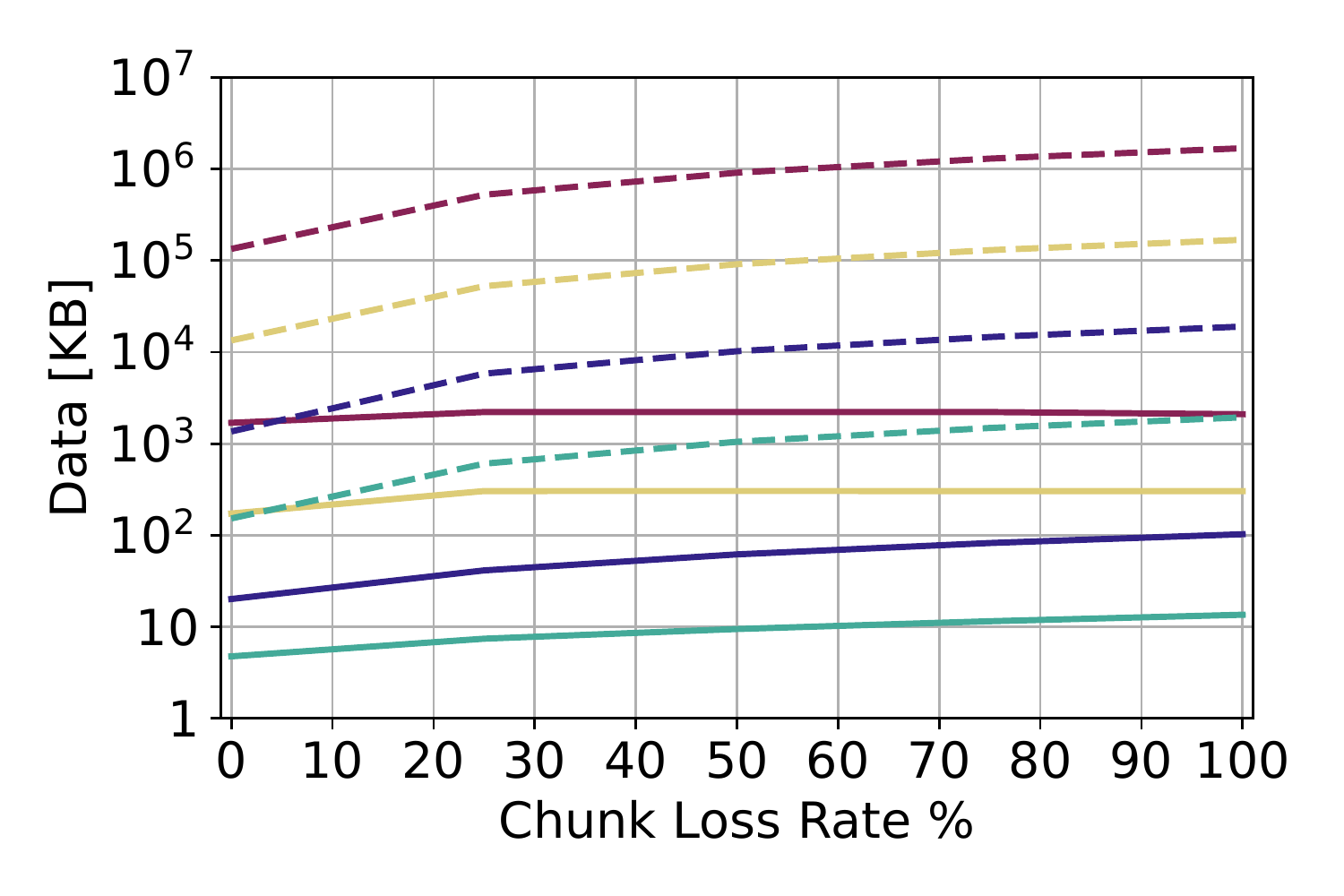}
		\caption{Data transmitted, $n$=17, CL}
		\label{fig:net-bandwidth-cl-86}
	\end{subfigure}
	\begin{subfigure}[t]{0.245\textwidth}
		\centering
		\includegraphics[width=\columnwidth,trim={1.02cm 1cm 0cm 0}]{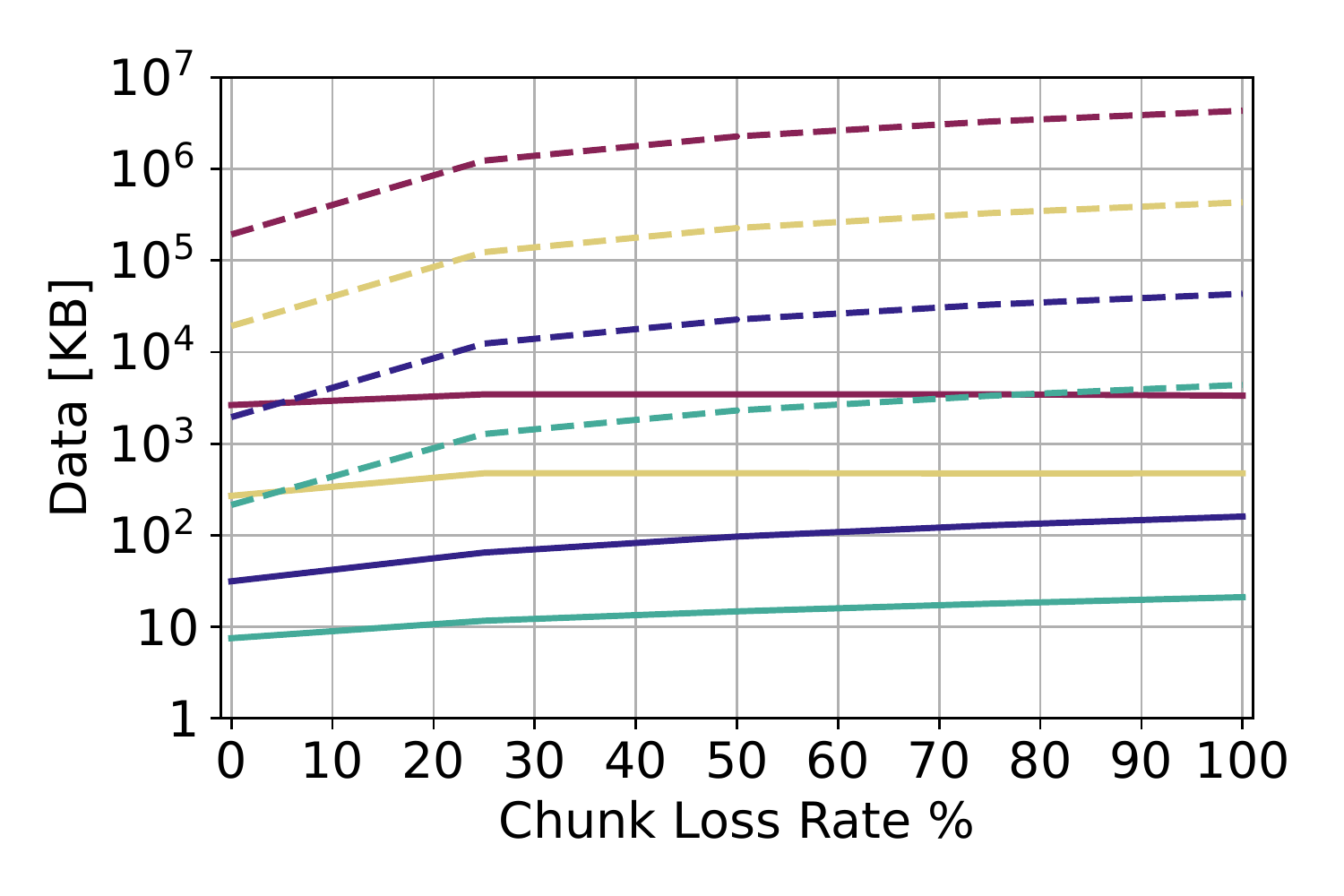}
		\caption{Data transmitted, $n$=26, CL}
		\label{fig:net-bandwidth-cl-f9}
	\end{subfigure}
	\begin{subfigure}[t]{0.245\textwidth}
		\centering
		\includegraphics[width=\columnwidth,trim={1.02cm 1cm 0cm 0}]{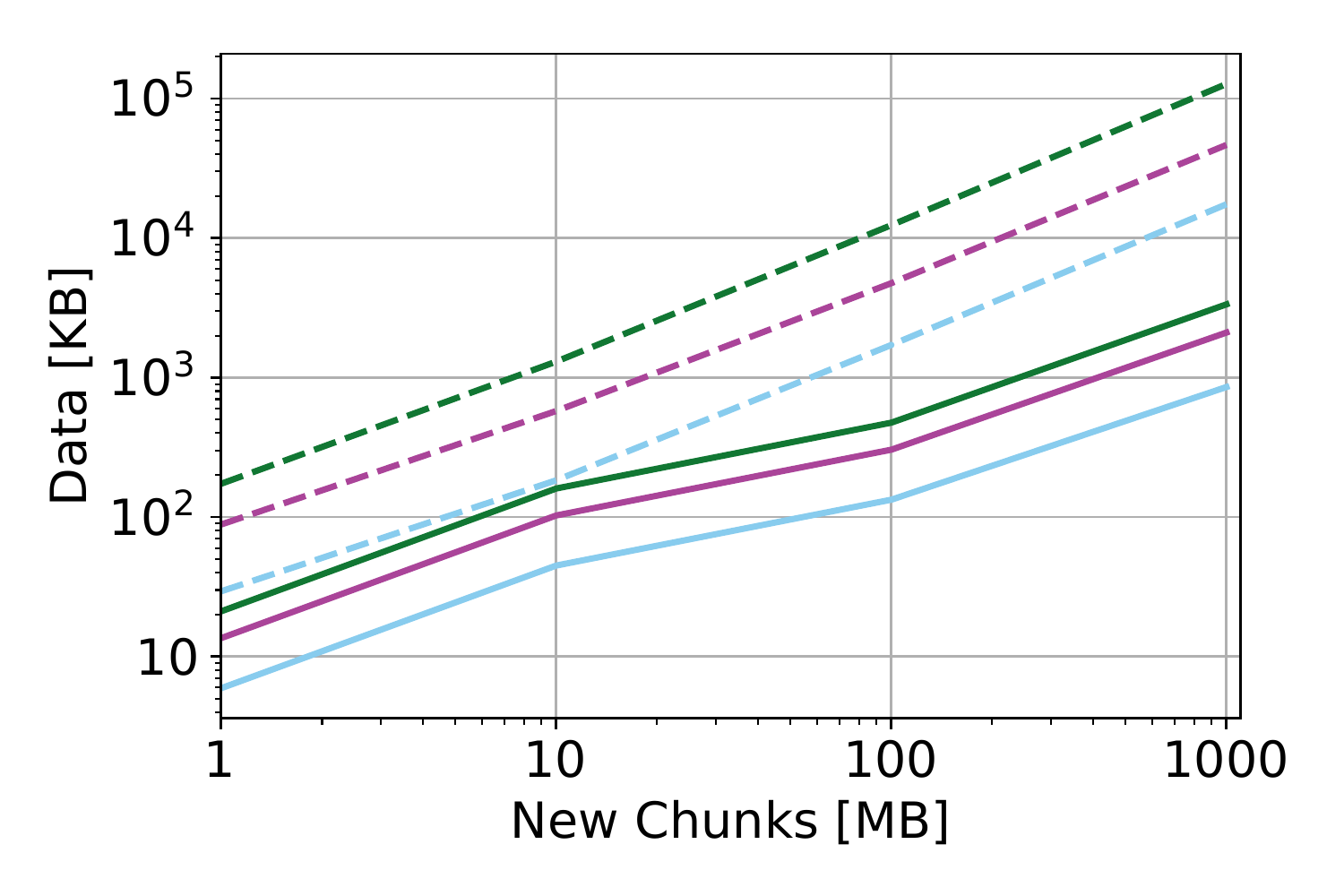}
		\caption{Data transmitted, CA}
		\label{fig:net-bandwidth-ca-all}
	\end{subfigure}

	\begin{subfigure}[t]{0.245\textwidth}
		\centering
		\includegraphics[width=\columnwidth,trim={1.02cm 1cm 0cm 0}]{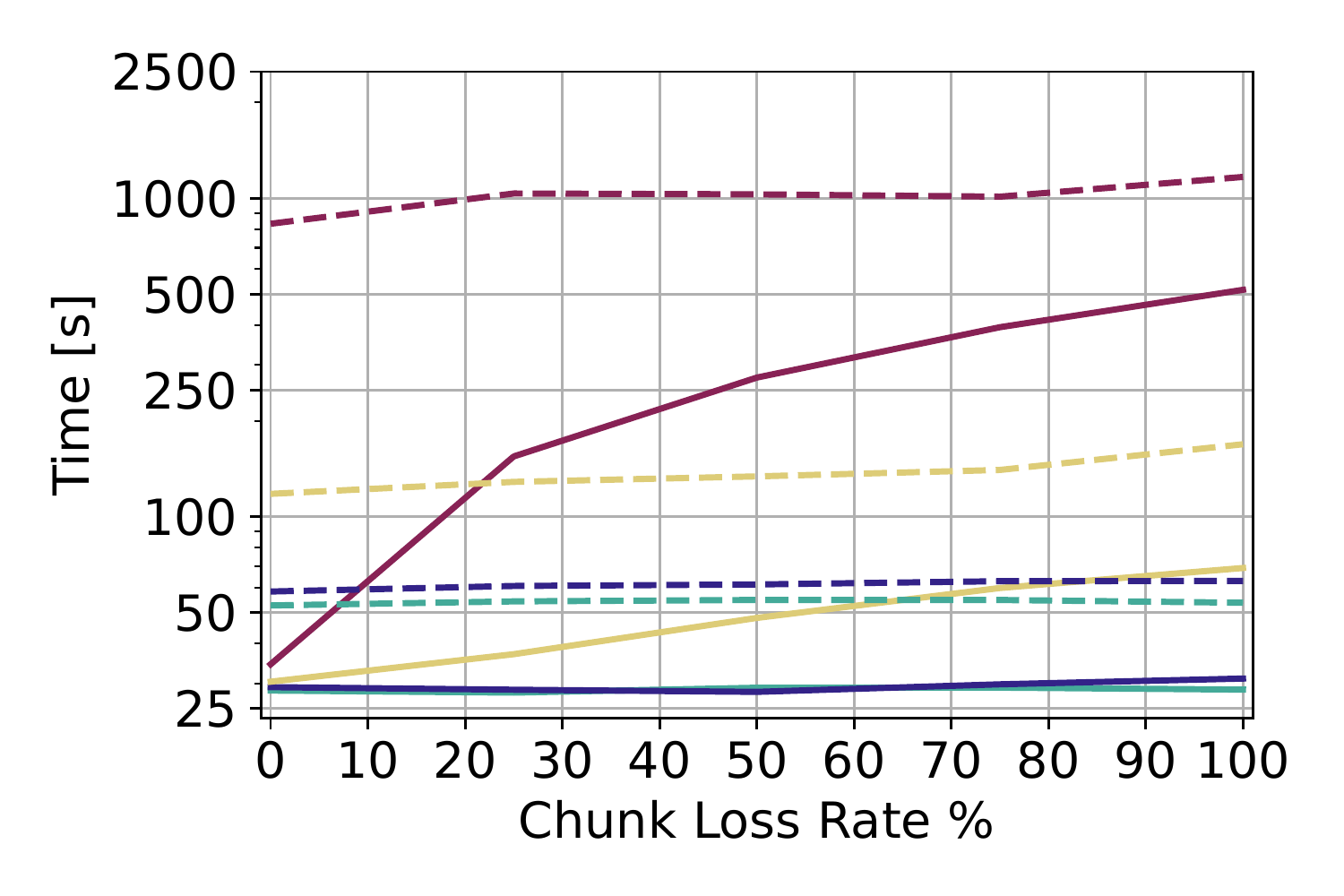}
		\caption{Duration, $n$=8, CL}
		\label{fig:net-duration-cl-97}
	\end{subfigure}
	\begin{subfigure}[t]{0.245\textwidth}
		\centering
		\includegraphics[width=\columnwidth,trim={1.02cm 1cm 0cm 0}]{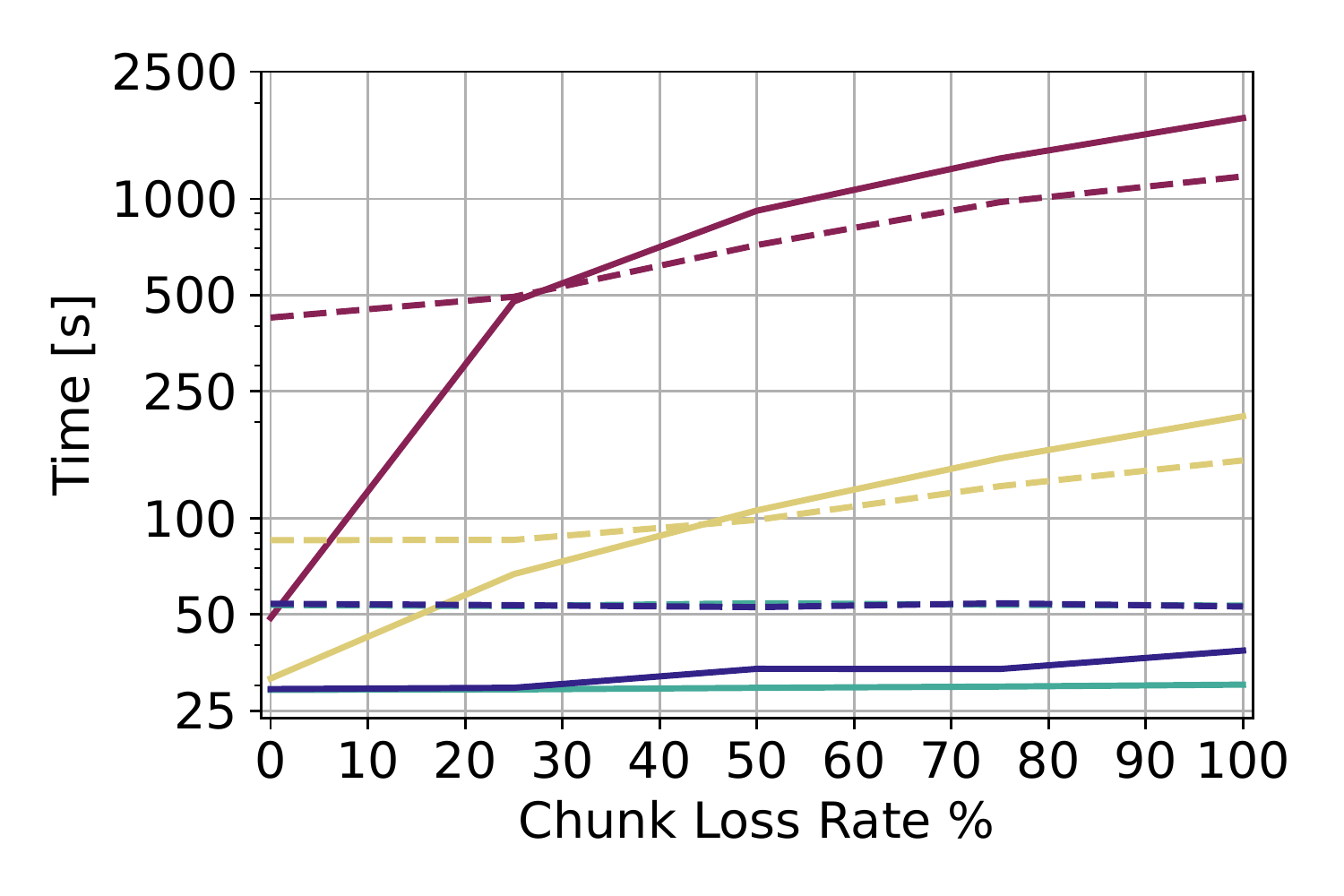}
		\caption{Duration, $n$=17, CL}
		\label{fig:net-duration-cl-86}
	\end{subfigure}
	\begin{subfigure}[t]{0.245\textwidth}
		\centering
		\includegraphics[width=\columnwidth,trim={1.02cm 1cm 0cm 0}]{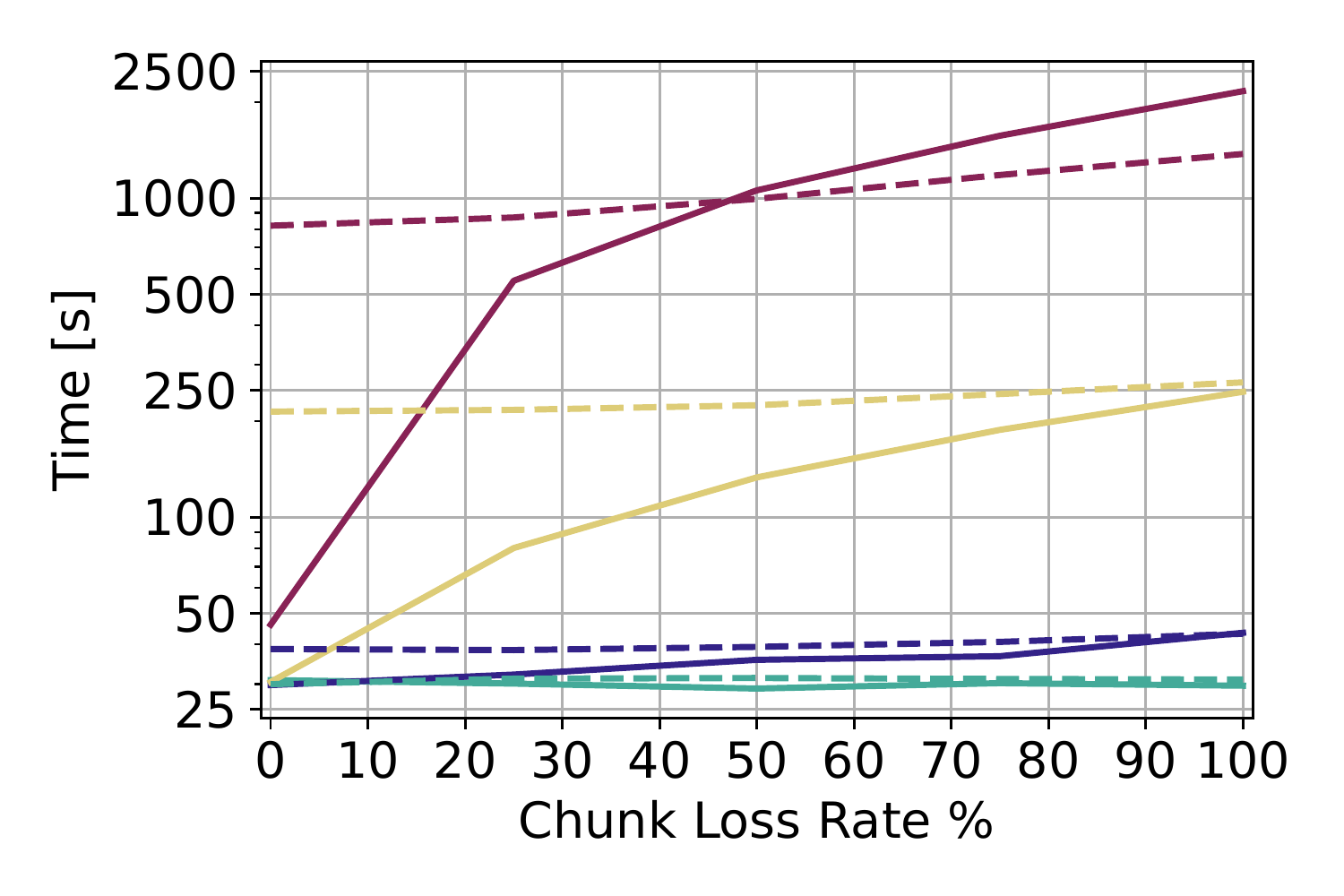}
		\caption{Duration, $n$=26, CL}
		\label{fig:net-duration-cl-f9}
	\end{subfigure}
	\begin{subfigure}[t]{0.245\textwidth}
		\centering
		\includegraphics[width=\columnwidth,trim={1.02cm 1cm 0cm 0}]{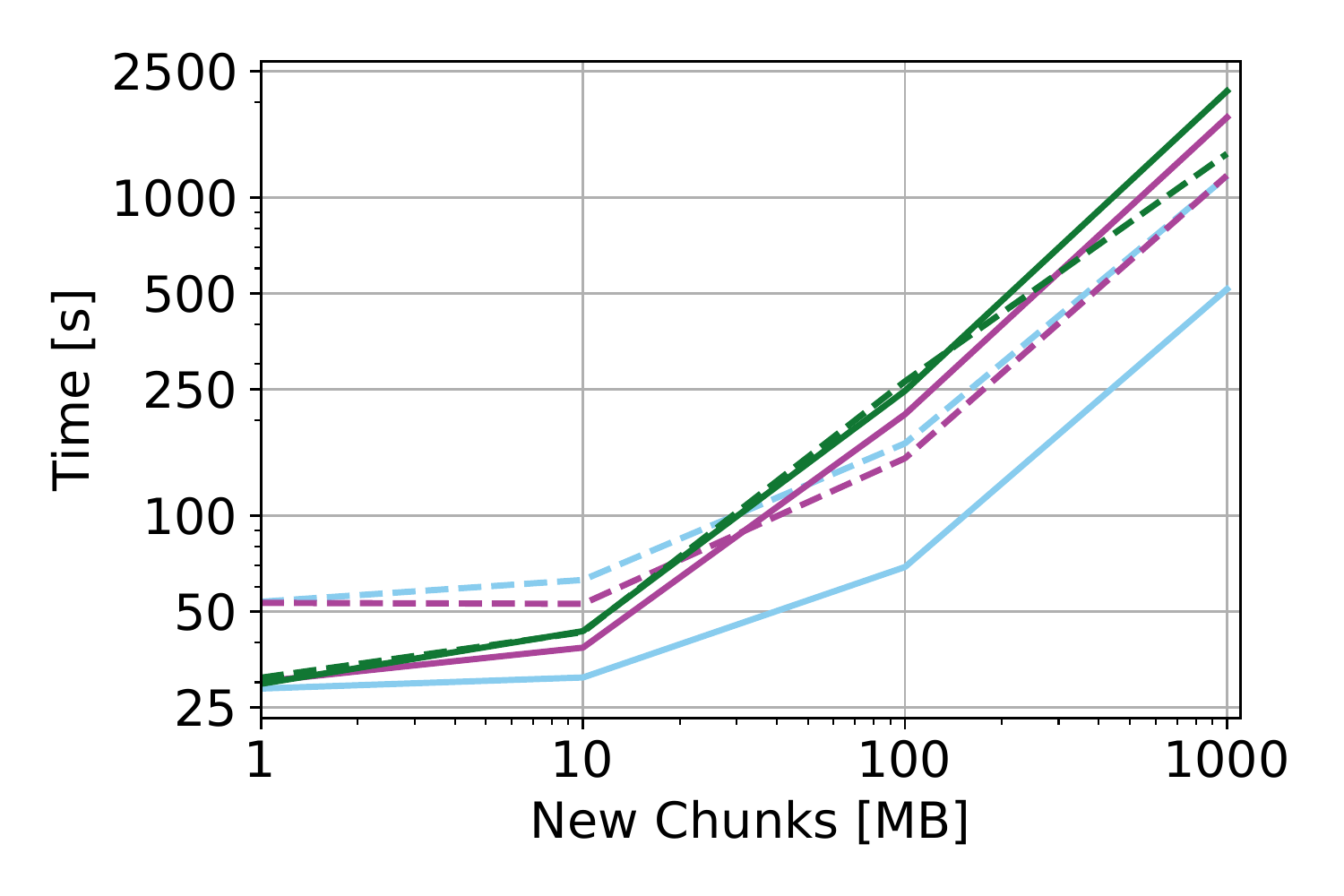}
		\caption{Duration, CA}
		\label{fig:net-duration-ca-all}
	\end{subfigure}

	\caption{Comparison between \protocolname and \pullsync when synchronizing peer neighborhoods.
	We vary the neighborhood storage sizes between 1, 10, 100, and 1000 MB.
	The peer neighborhood sizes $n=\{8, 17, 26\}$.
	Plots~(a,~b,~c) shows the data transmitted as we vary the chunk-loss rate (CL).
	Plots~(e,~f,~g) shows the duration as we vary the chunk-loss rate (CL).
	Plot~(d) shows the data transmitted as we vary the amount of new chunk data added (CA).
	Plot~(h) shows the duration as we vary the amount of new chunk data added (CA).
	The duration includes the time to transmit synchronization metadata and uploading chunks.}
	\label{fig:snarl-cost-effective}
\end{figure*}

In the CL scenario, peers initially store between $1$~MB and $1000$~MB of chunk data.
We then varied the amount of chunk loss from $0$~\% to $100$~\%.
\Cref{fig:snarl-cost-effective} (a,~b,~c) shows the \emph{metadata transmitted} to synchronize a neighborhood of sizes $8, 17$, and $26$, respectively.
\protocolname (solid lines) transmits 2-3 orders of magnitude less metadata than \pullsync (dashed lines).
The results are consistent for all neighborhood storage sizes.

\Cref{fig:snarl-cost-effective} (e,~f,~g) shows the \emph{time to synchronize} a neighborhood of sizes $8, 17$, and $26$, respectively.
The synchronization time includes downloading the synchronization metadata and uploading the chunks.
\protocolname is always faster than \pullsync for small storage sizes (1, 10~MB) and the 8-peer neighborhood.
For the 17- and 26-peer neighborhoods and larger storage sizes (100, 1000~MB), \pullsync is slightly faster when there is more than 30~\% and 50~\% chunk loss.
\pullsync is faster in these cases despite transmitting more metadata because of \pullsync's gossip optimization, explained in \Cref{sec:consistent-pullsync}.
Our experiments were performed on cluster nodes connected via unsaturated 10~Gbit/s links.
We conjecture that the gossip optimization will be less effective in an Internet environment where peers are connected over lower-bandwidth links.

In the CA scenario, we add $1$~MB to $1000$~MB of new chunk data to a peer and observe the synchronization behavior.
The \emph{metadata transmitted} to synchronize a neighborhood of sizes $8, 17$, and $26$ are shown in \Cref{fig:net-bandwidth-ca-all}.
As expected, the amount of metadata is linear with the new chunk data uploaded for \pullsync.
However, \protocolname has a sublinear relationship, which becomes more pronounced for larger uploads.
Overall, \protocolname transmits 1-1.5 orders of magnitude less data than \pullsync.

\Cref{fig:net-duration-ca-all} shows the \emph{time to synchronize} a neighborhood of sizes $8, 17$, and $26$.
\protocolname is always faster than \pullsync for 10~MB of new chunks.
However, due to the gossip optimization, \pullsync has a slight advantage when there are 100~MB or more new chunks.

We summarize the same results in \Cref{tab:bandwidthreduction} in the form of average \emph{bandwidth savings} and \emph{speedup}.

\begin{table*}[htbp]
	\renewcommand{\arraystretch}{1.05}
	\caption{Average bandwidth savings (KB) and speedup (ms) with \protocolname vs \pullsync. RSE is the relative standard error.}
	\label{tab:bandwidthreduction}
	\centering
	\footnotesize
	\begin{tabular}{ r@{\wx} | r@{\wx} | r@{\wx}r@{\wx}r@{\wx} | r@{\wx}r@{\wx}r@{\wx} | r@{\wx}r@{\wx}r@{\wx} | r@{\wx}r@{\wx}r@{\wx} }
		\toprule
		             &         & \multicolumn{3}{c|}{CL: Bandwidth Savings} & \multicolumn{3}{c|}{CA: Bandwidth Savings} & \multicolumn{3}{c|}{CL: Speedup} & \multicolumn{3}{c}{CA: Speedup} \\
		\mbox{Peers} & Size    & Avg~(KB) & Avg~(\%) & RSE~(\%)             & Avg~(KB) & Avg~(\%) & RSE~(\%)             & Avg~(ms) & Avg~(\%) & RSE~(\%)   & Avg~(ms) & Avg~(\%) & RSE~(\%)  \\
		\midrule
        8            &    1 MB &      308 &     98.7 & 0.4                  &       23 &     79.8 & 0.6                  &    25400 &     47.0 & 1.2        &    25069 &     46.6 & 1.4 \\
		             &   10 MB &     2981 &     99.1 & 0.1                  &      139 &     75.5 & 0.4                  &    31821 &     52.0 & 1.5        &    31859 &     50.6 & 2.2 \\
		             &  100 MB &    29849 &     99.6 & 0.1                  &     1582 &     92.2 & 0.1                  &    89170 &     64.6 & 1.9        &    99949 &     59.1 & 2.1 \\
		             & 1000 MB &   298957 &     99.7 & 0.0                  &    16630 &     95.1 & 0.3                  &   741126 &     73.0 & 1.5        &   651543 &     55.8 & 1.5 \\
		\midrule
		17           &    1 MB &     1038 &     99.1 & 0.4                  &       75 &     84.7 & 0.2                  &    24056 &     44.9 & 1.2        &    23150 &     43.4 & 1.0 \\
		             &   10 MB &    10166 &     99.4 & 0.1                  &      474 &     82.2 & 0.1                  &    20513 &     38.4 & 2.3        &    14385 &     27.2 & 3.7 \\
		             &  100 MB &    91024 &     99.7 & 0.0                  &     4449 &     93.6 & 0.1                  &    -3833 &     -3.5 & 1.4        &   -56847 &    -37.5 & 0.6 \\
		             & 1000 MB &   909988 &     99.8 & 0.0                  &    44463 &     95.5 & 0.0                  &  -156825 &    -20.7 & 1.5        &  -614137 &    -52.3 & 1.2 \\
		\midrule
		26           &    1 MB &     2293 &     99.4 & 0.3                  &      152 &     87.8 & 0.6                  &      955 &      3.1 & 1.3        &     1354 &      4.4 & 1.6 \\
		             &   10 MB &    22619 &     99.6 & 0.1                  &     1142 &     87.7 & 0.2                  &     4451 &     11.1 & 2.5        &     -307 &     -0.7 & 3.2 \\
		             &  100 MB &   226274 &     99.8 & 0.1                  &    11907 &     96.2 & 0.1                  &    96969 &     41.7 & 1.2        &    17451 &      6.6 & 1.3 \\
		             & 1000 MB &  2265603 &     99.9 & 0.0                  &   123972 &     97.4 & 0.1                  &   -29329 &     -2.8 & 1.6        &  -788973 &    -57.3 & 1.8 \\
		\bottomrule
	\end{tabular}
\end{table*}

\begin{figure*}[hbt]
	\centering
	\includegraphics[width=\textwidth]{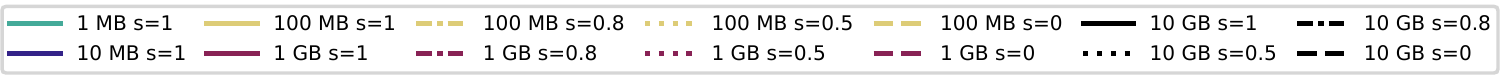}
	\begin{subfigure}[t]{0.320\textwidth}
		\centering
		\includegraphics[width=\columnwidth,trim={0cm 0.5cm 0cm 0}]{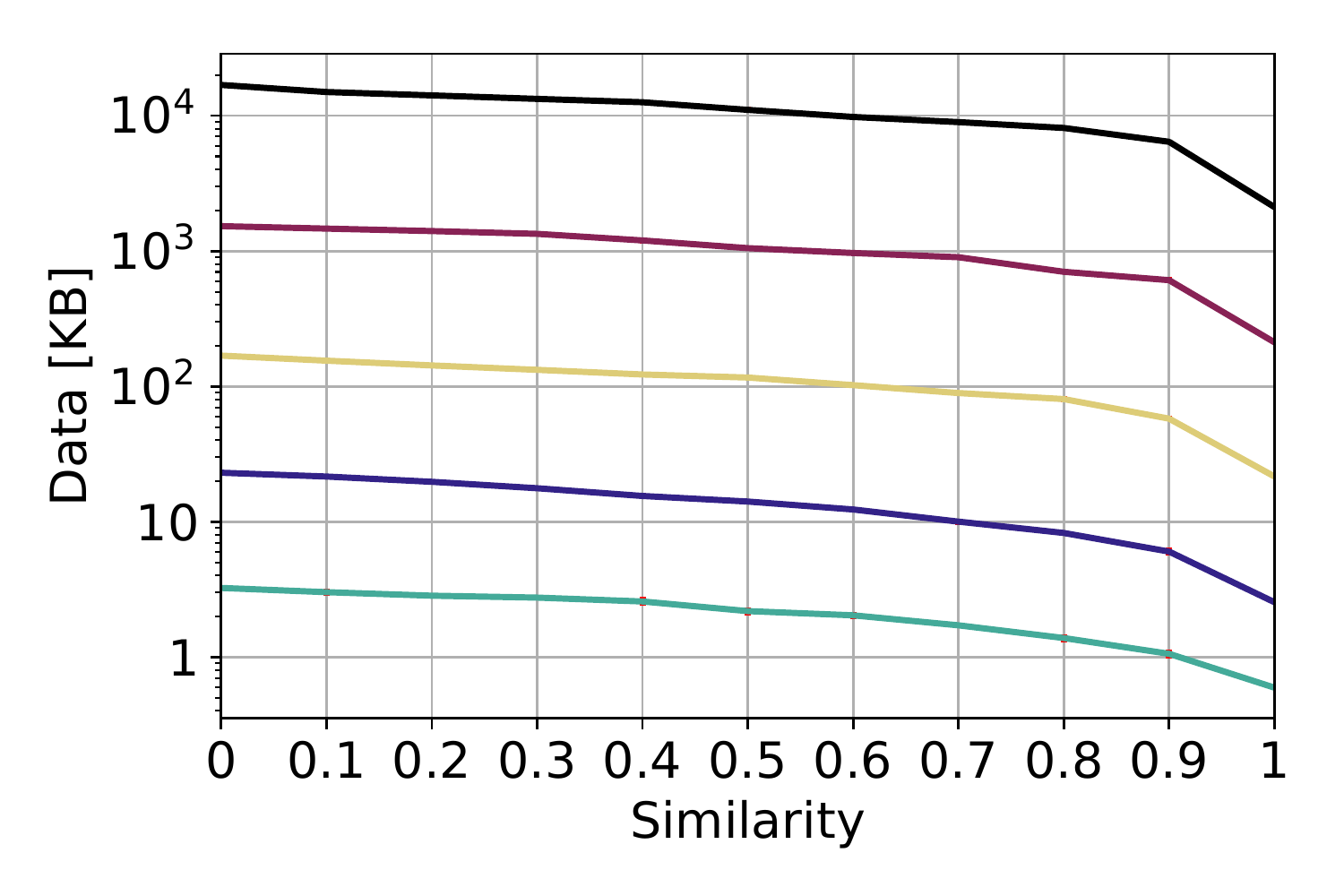}
		\caption{Data transmitted}
		\label{fig:similarity-bandwidth}
	\end{subfigure}
	\begin{subfigure}[t]{0.320\textwidth}
		\centering
		\includegraphics[width=\columnwidth,trim={0cm 0.5cm 0cm 0}]{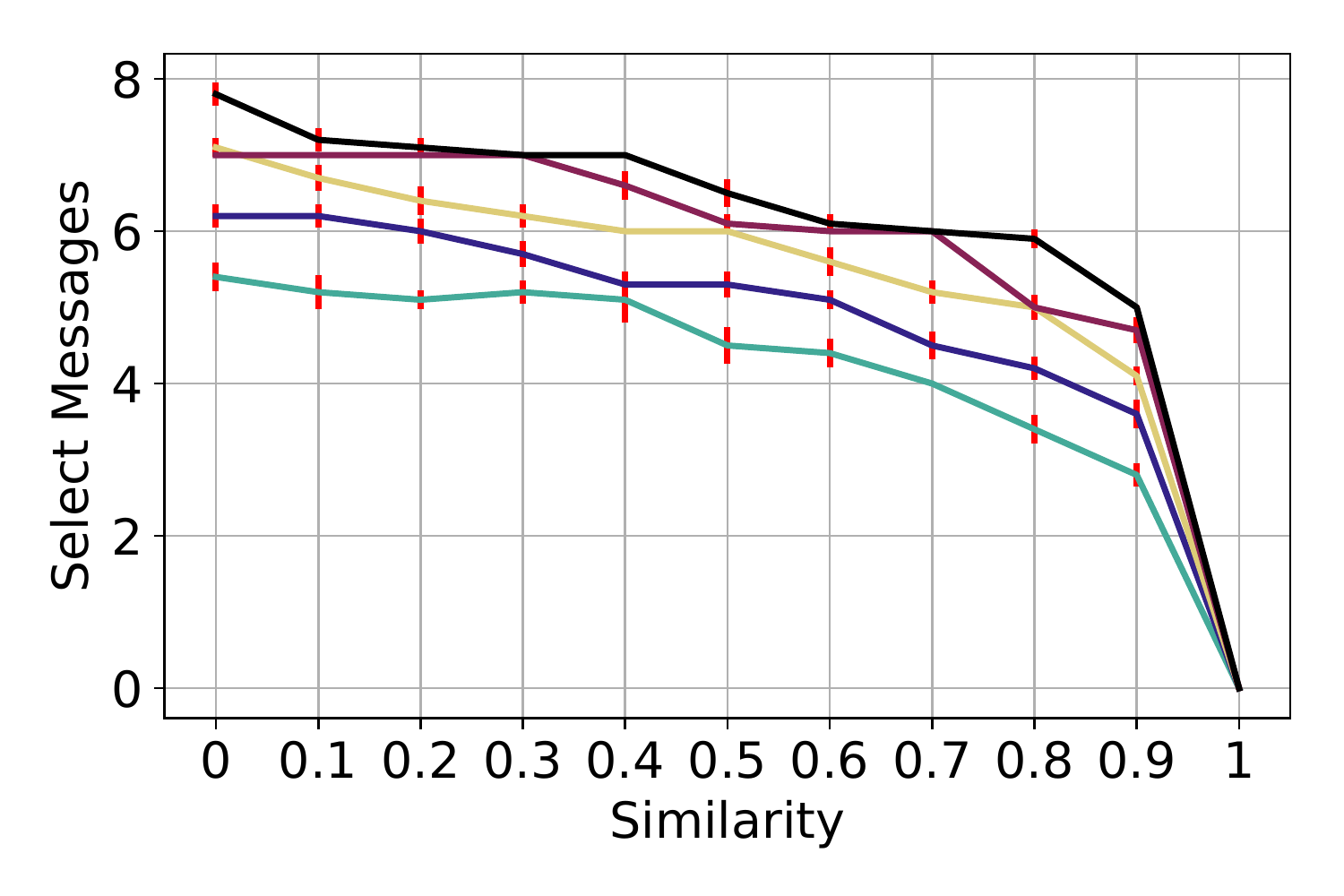}
		\caption{Select messages}
		\label{fig:similarity-select}
	\end{subfigure}
	\begin{subfigure}[t]{0.320\textwidth}
		\centering
		\includegraphics[width=\columnwidth,trim={0cm 0.5cm 0cm 0}]{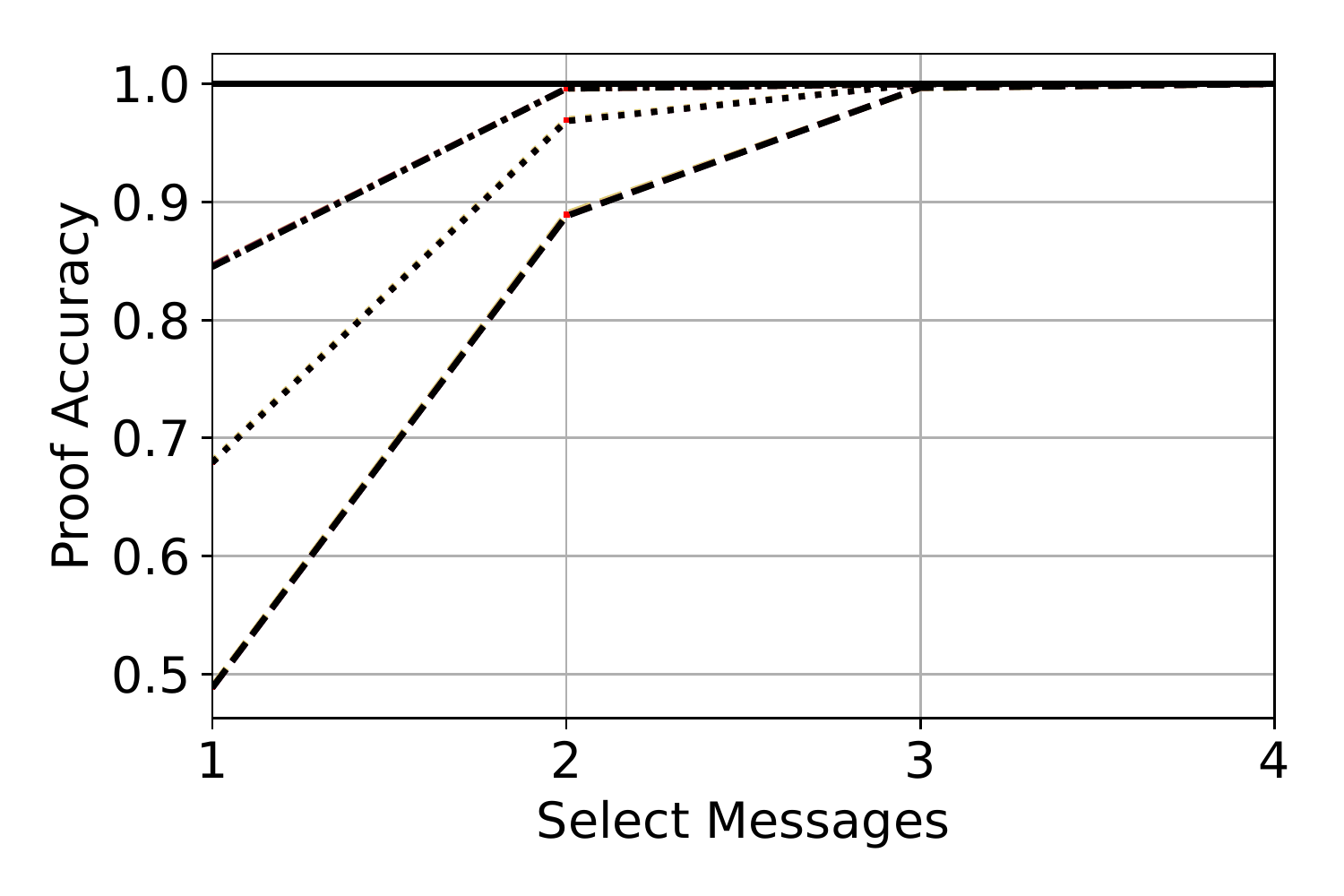}
		\caption{Proof accuracy}
		\label{fig:similarity-accuracy}
	\end{subfigure}
	\caption{\protocolname's performance with varying degrees of similarity $s = [0, 1]$.
    The labels define each peer's initial storage before the synchronization process begins.
	(a) Synchronization data for two peers.
	(b) Number of Select messages needed for two peers to fully synchronize.
	(c) Proof accuracy as a function of iterative Select messages.
	Vertical lines show the standard error.}
	\label{fig:sup-cost-effective}
\end{figure*}

\begin{figure*}[ht]
	\centering
	\begin{subfigure}[t]{0.325\textwidth}
		\centering
        \includegraphics[width=\columnwidth,trim={0cm 0.5cm 0cm 0}]{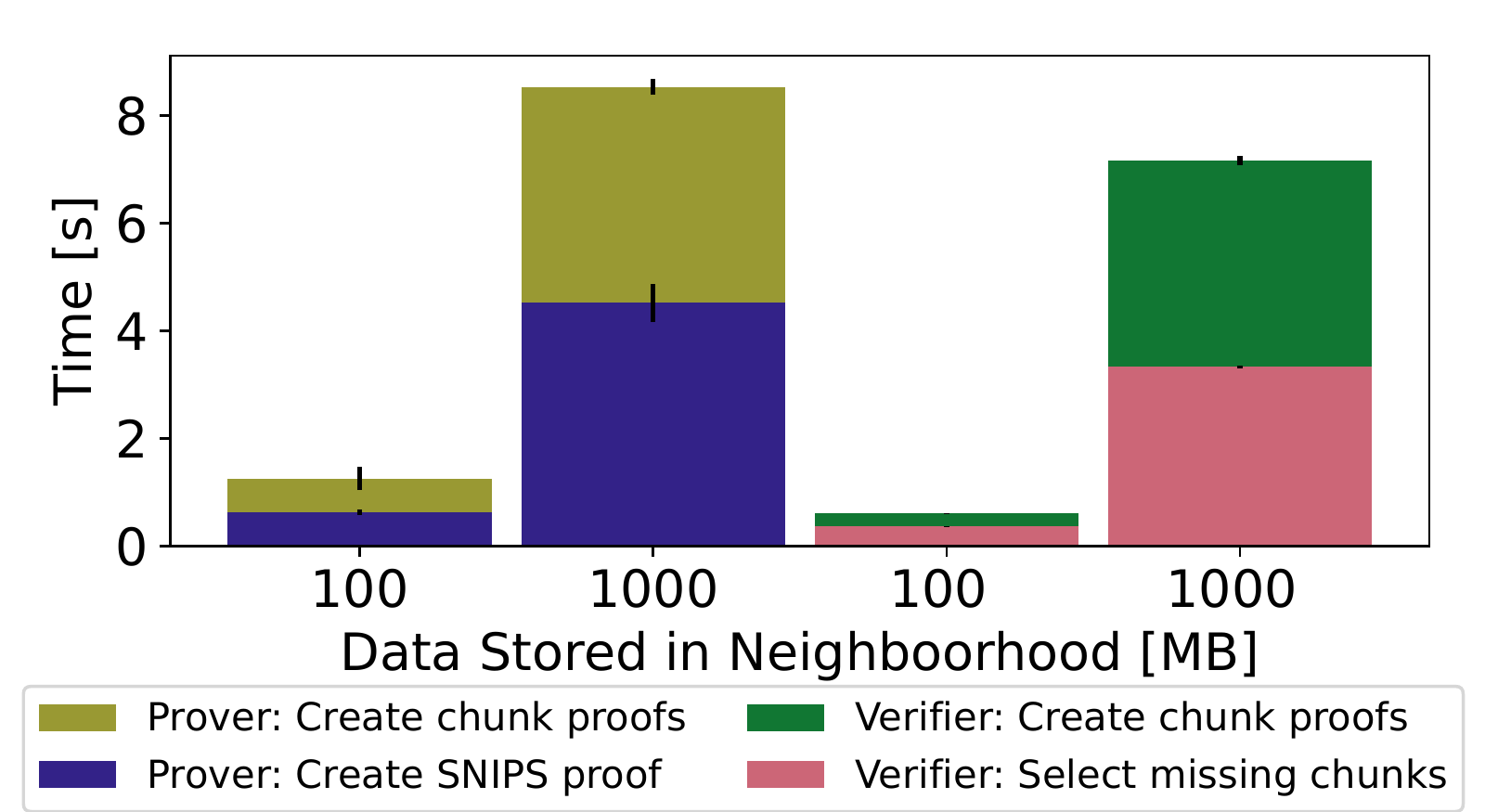}
        \caption{Time to create and select}
        \label{fig:eval-computation}
	\end{subfigure}
	\begin{subfigure}[t]{0.325\textwidth}
		\centering
        \includegraphics[width=\columnwidth,trim={0cm 0.5cm 0cm 0}]{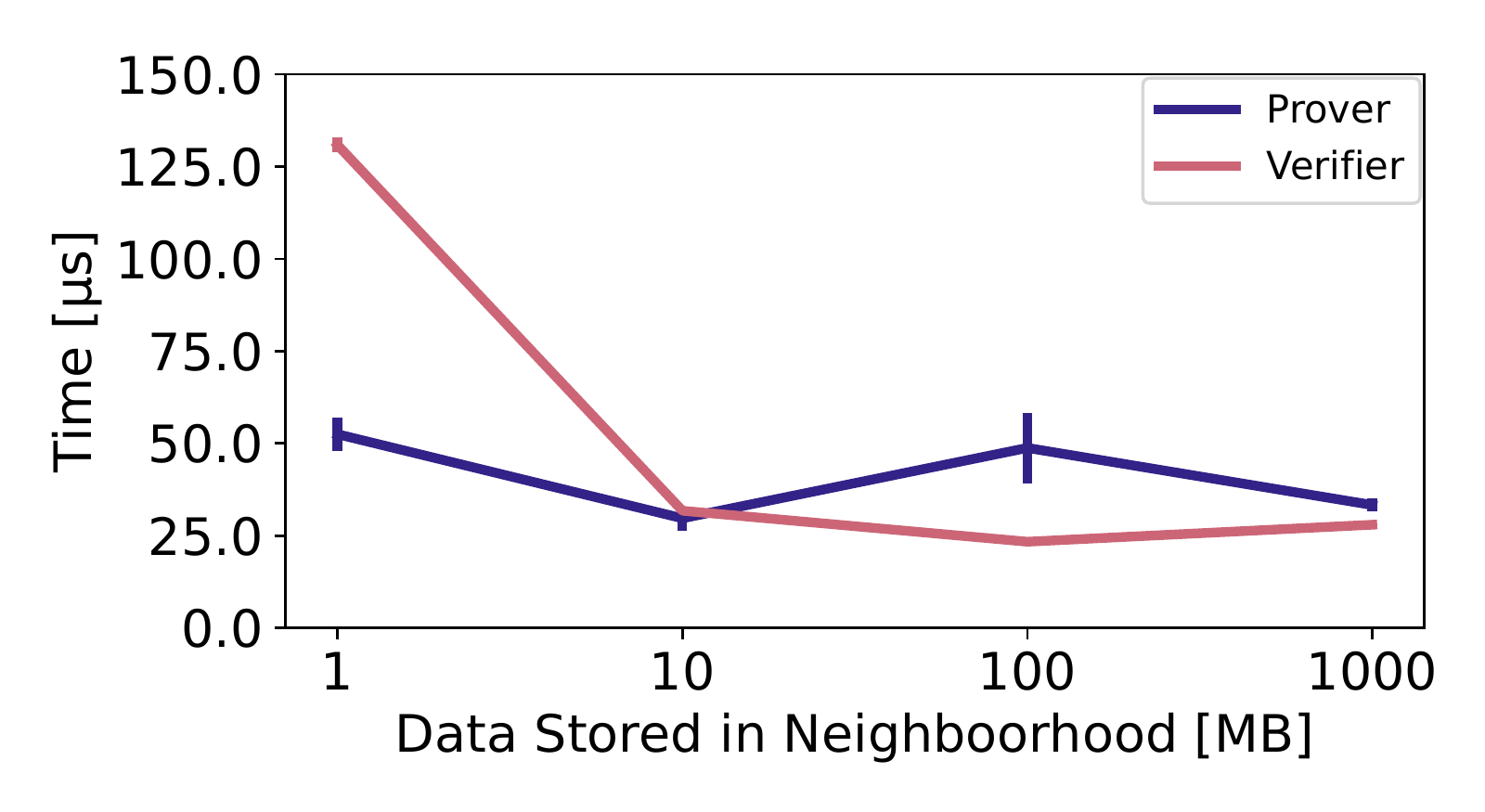}
        \caption{Time per chunk to create and select}
        \label{fig:eval-computation-per-chunk}
	\end{subfigure}
	\begin{subfigure}[t]{0.325\textwidth}
		\centering
        \includegraphics[width=\columnwidth,trim={0cm 0.5cm 0cm 0}]{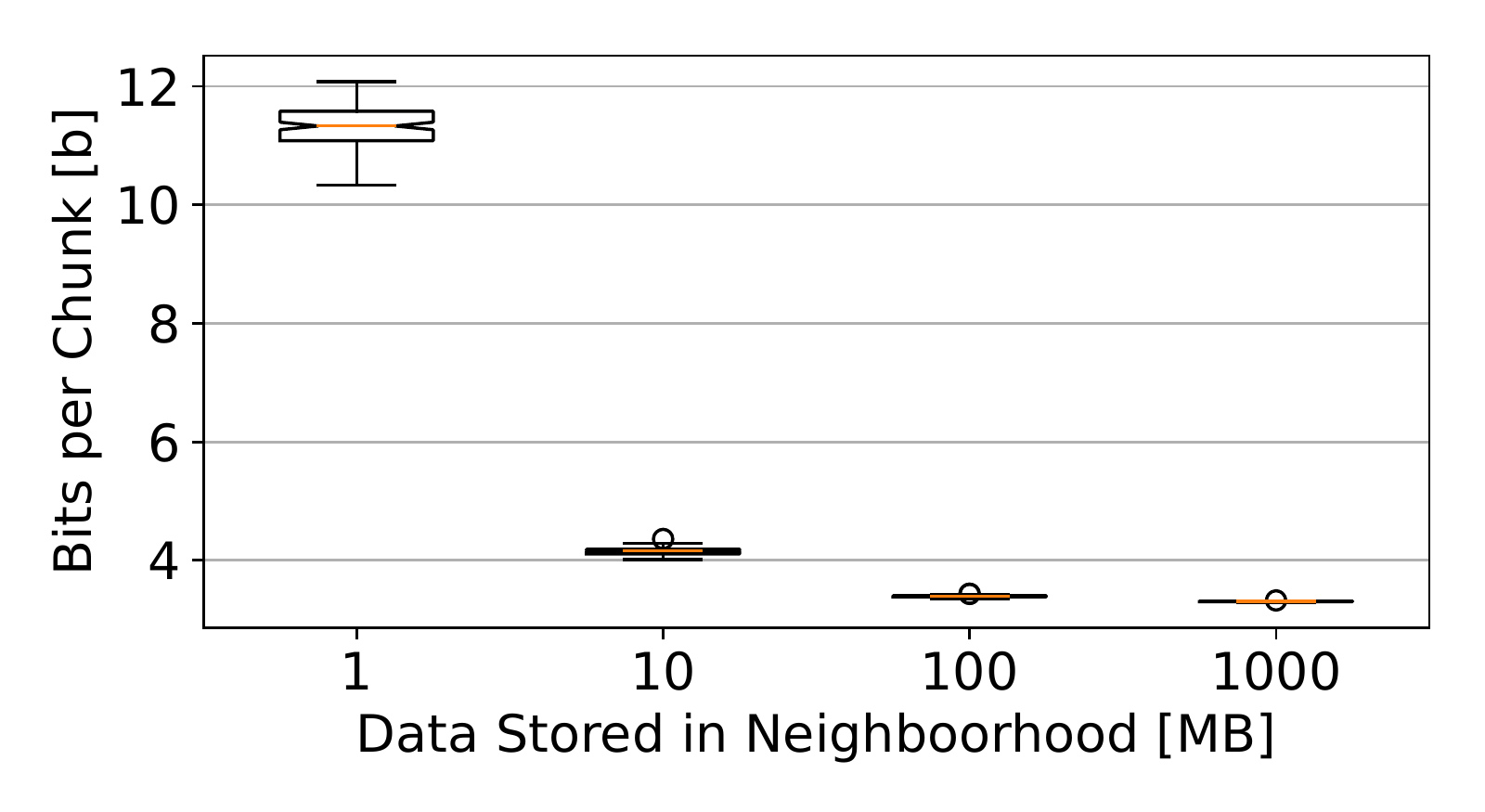}
        \caption{Per chunk overheads}
        \label{fig:eval-bits-per-chunk}
	\end{subfigure}
	\caption{Space and time overheads.
	(a)~Time spent creating and selecting missing chunks from a \protocolname proof.
	(b)~Time spent per chunk creating and selecting missing chunks from a \protocolname proof.
	(c)~Per chunk overhead for a \protocolname proof for different neighborhood storage sizes (bits per chunk).
    Vertical lines represent error bars.}
\end{figure*}

\subsection{\protocolname With Varying Degree of Similarity}
\label{sec:eval-similarity}

\noindent
Recall the metrics \emph{similarity}, $\vert A \cap B \vert / \min(\vert A \vert, \vert B \vert)$ and \emph{proof accuracy}, $|M_{i}| / |M_{p}|$, defined in \Cref{sec:sysmodel}.
We now simulate \protocolname's performance when similarity $< 1$, meaning both peers have chunks that the other does not.
As \Cref{sec:collision} explains, this complicates the synchronization as querying with chunk proofs not part of the original proof may cause collisions.

We initialize the peers with the same amount of chunks for all simulations.
Depending on the similarity setting, the total amount of chunks a peer is storing after concluding synchronization will range
 from the initial amount when similarity is 1 (chunks are fully overlapping)
 to $2\times$ the initial amount when similarity is 0 (chunks are fully disjoint).

We first simulate the transmitted synchronization metadata.
As expected, the transmitted metadata per peer when similarity is 1 in~\Cref{fig:similarity-bandwidth} closely matches the transmitted metadata when chunk loss is 0~\% in~\Cref{fig:snarl-cost-effective}.
For all storage sizes, the transmitted metadata increases by slightly less than one order of magnitude when similarity decreases from fully overlapping~(1) to fully disjoint~(0).
The growth in transmitted metadata comes from the need for additional sequential Prove and Select phases until the peers have fully synchronized.
For each round of Select, the peers become more synchronized, and their subsequent proofs become larger and more accurate.

\Cref{fig:similarity-select} shows the number of Select messages needed for two peers to synchronize fully.
As the peers initially have the same amount of chunks, when similarity is 1, peers are already synchronized, and no Select messages are needed.
The number of Select messages increases as the similarity decreases.

Finally, we evaluated the accuracy of the proofs, as shown in~\Cref{fig:similarity-accuracy}.
Interestingly, the amount of chunks does not impact the proof accuracy.
When similarity is~1, the proof accuracy is always~1, since all missing chunks are identified with a single Select message.
Moreover, the proof accuracy improves with the number of Select messages.
This is because each proof contains more chunks than the previous, allowing the peer to identify more missing chunks.
The highest number of Select messages needed to fully synchronize was 4.

\begin{table}[htbp]
	\renewcommand{\arraystretch}{1.1}
	\caption{Time spent creating and selecting missing chunks from a \protocolname proof.}
	\label{tab:computational}
	\centering
	\footnotesize
	\begin{tabular}{ l@{\wx} | c@{\wx}c@{\wx} | c@{\wx}c@{\wx} }
		\toprule
		&  \multicolumn{2}{c|}{Prover} & \multicolumn{2}{c}{Verifier} \\
		\mbox{Operation} & 100 MB & 1000 MB & 100 MB & 1000 MB \\
		\midrule
		Chunk proofs & 622 ms &  4517 ms & 359 ms & 3327 ms \\
		Create & 628 ms &  4017 ms & & \\
		Select & & & 240 ms & 3838 ms \\
		\bottomrule
	\end{tabular}
\end{table}

\subsection{Computation Overhead}
\label{sec:eval-computation-overhead}

\noindent
We measured the time spent creating and selecting missing chunks (verifying) from a \protocolname proof by extracting the time from the logs obtained from running the experiment in~\Cref{sec:eval-realworld} on our cluster.
We isolate the time spent creating chunk proofs as this is common for both operations.
Our results are shown in~\Cref{fig:eval-computation} and~\Cref{tab:computational}.

The results show that the verifier spends slightly less time creating chunk proofs than the prover.
In addition, the time spent selecting missing chunks is slightly less than the time spent creating the proof.

In comparison, we found that the average time to answer one peer with a long list of chunk identifiers in Pullsync's Offer phase amounts to 73.2 seconds for 100~MB of chunks and 728.6 seconds for 1000~MB of chunks.
The Offer phase is executed partially in parallel and thus further study is needed to determine the exact time spent in sequential operations.

Next, we measured how the time spent correlates with the size of the proof.
\Cref{fig:eval-computation-per-chunk} shows that only tens of microseconds are necessary for either operation.
As the size of the proof increases, the time spent per chunk seems to stabilize.
The time per chunk is slightly higher for 1~MB proofs due to the constant overhead for signing the proof.
This overhead is amortized over many chunks for larger proofs, thus lowering the per-chunk time.

\subsection{Storage Overhead}
\label{sec:eval-storage-overhead}

\noindent
We extracted the size of the \protocolname proofs from the experiments presented in~\Cref{sec:eval-realworld}.
As the size of the proof is proportional to the number of chunks, we measured the storage overhead in terms of bits per chunk.
Our results are shown in~\Cref{fig:eval-bits-per-chunk}.
The overhead approaches 3.3 bits per chunk for a proof containing 1000~MB of chunks.
As in~\Cref{sec:eval-computation-overhead}, the overhead for 1~MB of chunks is dominated by constant overheads.
In this case, the higher number of bits per chunk for 1~MB of chunks is due to the digital signature, the nonce, and the $[ \varstart, \varend ]$ range.
However, for larger storage sizes, this overhead is amortized.

\section{Related Work}
\label{sec:related-works}

\noindent
We contrast our work with two main categories of related work: PoS constructions and data synchronization protocols.

\subsection{Bloom Filter Variants and Set Reconciliation}
\noindent
A Bloom filter~\cite{bloom1970space} is a probabilistic data structure for dynamic sets that supports a similar membership query as our PoS-like construction.
The Bloom filter uses $k$ hash functions to map each element to $k$ bits in a bit array.
When querying a Bloom filter, the results will be $k$ array positions.
If either of the $k$ array positions is $0$, then the element is \emph{definitely not in the set}.
Otherwise, we say that the element \emph{might be in the set}, as the membership query suffers from false positives.
The false positive rate of a Bloom filter can be improved by increasing the number of bits used per element.

A crucial difference between the Bloom filter and our PoS-like construction is that our membership query always returns an index value index value $[0, N]$ that can be used to identify missing elements.
As pointed out in previous works~\cite{eppstein2010straggler}, it is not obvious how to identify missing elements from the Bloom filters array positions, other than testing membership on all possible elements in the universe.

An \ac{IBF}~\cite{eppstein2010straggler} is an extension to the standard Bloom filter that allows a query to extract the elements in the set---if the set is small enough.
The \ac{IBF} embeds an XOR field in each cell which is calculated over all the keys in the cell.
Two \ac{IBF}s can attempt to use the XOR fields to identify the elements in their symmetric difference.

Set reconciliation using \ac{IBF} was suggested in~\cite{eppstein2011s}.
Their results show that the method can only identify the missing elements with high probability when the symmetric difference is less than 30~\%.
In comparison, we have shown that \protocolname can always identify the missing elements, even when the symmetric difference is 100~\% (the maximum).

A cuckoo filter~\cite{fan2014cuckoo} is an alternative to a Bloom~filter that supports deletion and approximate membership tests.
Elements are inserted into the cuckoo filter by hashing the element with two different hash functions and then mapping the hash value to at least one of the two resulting candidate buckets.

Recent work~\cite{luo2021mcfsyn} has proposed \emph{MCFSyn} for multi-party set reconciliation using marked cuckoo filters (MCF).
Each peer generates an MCF vector that is sent to a centralized participant.
The centralized participant then creates an overall MCF vector and distributes it to the peers.
The peers can then identify the missing elements by comparing their MCF vectors with the overall MCF vector.
The protocol requires that the set of peers are fixed for the duration of the protocol.
Both false positives (identifying a missing element as being in the set) and false negatives (identifying an element as missing from the set) are possible in the protocol, resulting in excessive communication and inaccurate reconciliation.
In contrast, \protocolname is a decentralized protocol that does not require a fixed set of peers and does not suffer false negatives.

\subsection{Proof of Storage Algorithms and Accumulators}
\label{sec:related-works-pos}
\noindent
\ac{PoS} algorithms~\cite{ateniese2007provable,juels2007pors,shacham2008compact} allow a peer to prove that it possesses a chunk of data without revealing the chunk itself.
Three actors are involved in a PoS protocol: a challenger, a prover, and a verifier.
As summarized in~\cite{yang2018lightweight}, the features, security, and performance of PoS algorithms vary greatly.

A related feature is \emph{public verifiability}, first introduced in~\cite{ateniese2007provable}.
With this feature, anyone can take on the verifier role and verify the proof.
Algorithms with this feature require that the original data owner generates and persists some metadata to be used by other peers to create storage challenges and to verify the proofs.
The metadata adds extra overhead, and it needs to be generated for each data chunk that should be publicly verifiable.
Should this metadata be lost, the original data owner must regenerate it.
Our PoS-like construction allows anyone to possess the original data set to verify the proof without needing additional metadata.

We mention briefly that some PoS schemes~\cite{damgaard2019proofs,curtmola2008mr} can detect colluding peers by encoding each replica differently or by relying on timing assumptions.
However, these reliances are unsuitable for decentralized storage systems.

An accumulator is a cryptographic construction representing a set of elements and allows the issuance of membership proofs without revealing the elements themselves~\cite{benaloh1993one}.
A universal accumulator~\cite{li2007universal} allows a prover to generate both membership and non-membership proofs.
Other variants include accumulators based on Bilinear maps~\cite{nguyen2005accumulators} and Merkle trees~\cite{baldimtsi2017accumulators}.
Unlike our PoS-like construction, accumulators can give definite answers to membership queries.
However, they would require a different witness for each chunk, which must be constantly updated, potentially overwhelming the system.

\subsection{rsync}
\noindent
A popular tool for data synchronization is \emph{rsync}.
The rsync algorithm~\cite{tridgell1999efficient} is designed to reduce the amount of data needed to transfer files between two peers.
Rsync achieves this by only transferring blocks not already at the destination and using a rolling checksum to detect changes within files.
However, rsync is unsuitable for decentralized storage systems as it is limited to synchronizing one peer at a time.

\section{Conclusion}
\label{sec:conclusion}

\noindent
This work presents \protocolname, a novel protocol for data synchronization in decentralized storage systems.
Having efficient data synchronization that can run frequently and respond to network churn is paramount for keeping sufficient redundancy levels in the storage system.
\protocolname features a PoS-like construction for creating storage proofs that support membership queries.
Peers exchange storage proofs with neighboring peers and query the proofs to identify missing chunks and subsequently request them.
The proofs typically require only a few bits per chunk.

Our contribution includes rigorous experiments on a real-world cluster to show that \protocolname is a practical protocol.
In addition, we have demonstrated the correctness of \protocolname by simulating the performance under worst-case scenarios.

We compare \protocolname to \pullsync, the state-of-the-art protocol for data synchronization in Ethereum Swarm.
Our results show that \pullsync is vulnerable to inconsistencies, does not provide storage guarantees, and uses up to three orders of magnitude more synchronization data than \protocolname.

The potential impact of \protocolname on other decentralized storage systems seems promising but is yet to be fully explored.
We believe there are other uses for the PoS-like construction, but we leave them for future work.

\bibliographystyle{IEEEtran}
\bibliography{references}

\end{document}